\documentclass{emulateapj}

\newcommand \microjy{$\mu$Jy}
\newcommand \plagas{PLGs}
\newcommand \plaga{PLG}

\def\gsim{\mathrel{\rlap{\lower4pt\hbox{\hskip1pt$\sim$}} \raise1pt\hbox{$>$}}}
\def\lsim{\mathrel{\rlap{\lower4pt\hbox{\hskip1pt$\sim$}} \raise1pt\hbox{$<$}}}

\usepackage{lscape}

\slugcomment{Accepted for publication in the Astrophysical Journal; 2008 June 26}
\shorttitle{\textit{Spitzer}--Selected AGN}
\shortauthors{DONLEY ET~AL}

\begin{document}

\title{\textit{Spitzer's} Contribution to the AGN Population}

\author{J. L. Donley, \altaffilmark{1} G. H. Rieke, \altaffilmark{1}
P. G. P\'{e}rez-Gonz\'{a}lez, \altaffilmark{2,3} G. Barro\altaffilmark{2}}

\altaffiltext{1}{Steward Observatory, University of Arizona, 933 
North Cherry Avenue, Tucson, AZ 85721; jdonley@as.arizona.edu}
\altaffiltext{2}{Departamento de Astrof\'{\i}sica y CC. de la Atm\'osfera, Facultad de
CC. F\'{\i}sicas, Universidad Complutense de Madrid, 28040 Madrid,
Spain}
\altaffiltext{3}{Associate Astronomer at Steward Observatory, The University of Arizona}

\begin{abstract}

Infrared selection is a potentially powerful way to identify heavily
obscured AGN missed in even the deepest X-ray surveys.  Using a
24~\micron-selected sample in GOODS-S, we test the reliability and
completeness of three infrared AGN selection methods: (1) IRAC
color-color selection, (2) IRAC power-law selection, and (3) IR-excess
selection; we also evaluate a number of infrared excess approaches.
We find that the vast majority of non-power-law IRAC color-selected
AGN candidates in GOODS-S have colors consistent with those of
star-forming galaxies. Contamination by star-forming galaxies is most
prevalent at low 24~\micron\ flux densities ($\sim 100$~\microjy) and
high redshifts ($z\sim 2$), but the fraction of potential contaminants
is still high ($\sim 50\%$) at 500~\microjy, the highest flux density
probed reliably by our survey.  AGN candidates selected via a simple,
physically-motivated power-law criterion (\plagas), however, appear to
be reliable.  We confirm that the infrared excess methods successfully
identify a number of AGN, but we also find that such samples may be
significantly contaminated by star-forming galaxies.  Adding only the
secure \textit{Spitzer}-selected \plaga, color-selected, IR-excess,
and radio/IR-selected AGN candidates to the deepest X-ray--selected
AGN samples directly increases the number of known X-ray AGN (84) by
$54-77\%$, and implies an increase to the number of
24~\micron-detected AGN of $71-94\%$. Finally, we show that the
fraction of MIR sources dominated by an AGN decreases with decreasing
MIR flux density, but only down to $f_{\rm 24~\micron} =
300$~\microjy.  Below this limit, the AGN fraction levels out,
indicating that a non-negligible fraction ($\sim 10\%$) of faint
24~\micron\ sources (the majority of which are missed in the X-ray)
are powered not by star formation, but by the central engine.  The
fraction of all AGN (regardless of their MIR properties) exceeds 15\%
at all 24~\micron\ flux densities.

\end{abstract}


\section{Introduction}

Identifying complete and reliable samples of AGN has become a
necessity for extragalactic surveys, whether the goal be the selection
of AGN candidates or the removal of AGN ``contaminants''.  Only when
armed with complete samples of AGN will we be able to determine the
role of obscured accretion in the build-up of the present day black
hole mass function, or accurately characterize the star-formation
history of the universe.  Complete AGN samples are also required to
test proposed evolutionary theories in which black hole formation and
star-formation are intimately linked by merger and feedback processes
(e.g. Hopkins et al. 2006), ultimately producing the correlation
between black hole mass and bulge velocity dispersion (Ferrarese \&
Merritt 2000; Gebhardt et al. 2000). Unfortunately, the varied
luminosities, accretion rates, orientations, and intrinsic
obscurations of AGN prevent any one selection technique from reliably
identifying all of them.  For instance, while current UV, optical, and
X-ray surveys are capable of detecting unobscured AGN, they miss many
of the obscured AGN and nearly all of the Compton-thick AGN thought to
dominate AGN number counts at both low and high redshift (e.g. Gilli
et al. 2007; Daddi et al. 2007a,b).  Likewise, only 10-15\% of AGN are
radio-loud, making radio surveys relatively incomplete.

The Multiband Imaging Photometer (MIPS; Rieke et al. 2004) and
Infrared Array Camera (IRAC; Fazio et al. 2004) instruments aboard
\textit{Spitzer} have provided sensitive surveys in multiple mid-IR bands. 
Infrared selection with MIPS and IRAC data is being used widely to
select AGN candidates independently of their optical and/or X-ray
properties.  In addition to identifying AGN in fields with little or
no X-ray data, infrared selection criteria are capable of identifying
heavily obscured AGN missed in even the deepest X-ray fields
(e.g. Donley et al. 2007). As such, IR selection has the potential to
complement traditional AGN selection methods and to yield a more
complete census of AGN activity.

In this paper, we critically review the following infrared selection
criteria: (1) IRAC color-color selection, (2) IRAC power-law
selection, and (3) IR-excess selection.  The first selection method
employs color cuts in two representations of IRAC 4-band mid-infrared
(MIR) color-color space (Lacy et al. 2004; Stern et al. 2005), the
second identifies AGN whose IRAC SEDs are well-fit by a power-law
(Alonso-Herrero et al. 2006, Donley et al. 2007), and the third
selects red galaxies with large infrared to UV/optical flux ratios
(Daddi et al. 2007a, Dey et al. 2008, Fiore et al. 2008, Polletta et
al. 2008).  The first two criteria are based on the same principle:
the hot dust near an AGN's central engine reprocesses absorbed UV,
optical, and X-ray emission into short-wavelength MIR emission,
filling in the gap between the stellar emission that peaks near
1.5~\micron\ and the long-wavelength dust emission features that
dominate the SEDs of star-forming galaxies.  The color-color and
power-law selection criteria, however, differ in the range of mid-IR
characteristics they include as possible AGN indicators.  The third
selection method identifies sources in which heavy obscuration with
reemission in the infrared diminishes the optical emission and/or
enhances the infrared emission.

This paper utilizes improved spectral templates with a sample of
infrared color-selected, power-law galaxies (\plagas), and IR-excess
galaxies in the ultra-deep GOODS-S field to test the reliability and
completeness of these selection techniques over a wide range of sample
properties.  From this analysis, we then quantify the contribution of
these approaches plus \textit{Spitzer} identification of
radio-intermediate and radio-loud AGN to the X-ray--selected AGN
population.  The paper is organized as follows.  In \S2, we describe
the selection of the color-selected, \plaga, and IR-excess samples.
The construction of high-reliability photometric redshifts is
described in \S3, as are the overall redshift properties of the
sample.  In \S4, we briefly discuss the X-ray properties of our
MIPS-selected sample.  The infrared color selection criteria of Lacy
et al. (2004) and Stern et al. (2005) are discussed in \S5, where we
compare and contrast the two selection criteria, investigate the
behavior in color space of the star-forming templates, determine the
redshift and flux dependencies of the color selection techniques, and
investigate the properties of the most secure color-selected AGN
candidates.  In \S6, we discuss the
\plaga\ selection criteria, and in \S7 we investigate the 
IR-excess sources. Finally, in \S8, we discuss the overall statistics
of AGN revealed by IR-based methods compared with X-ray--selected
samples.  A summary is given in \S9.  Throughout the paper, we assume
the following cosmology: ($\Omega_{\rm m}$,$\Omega_{\rm
\Lambda},H_0$)=(0.3, 0.7, 72~km~s$^{-1}$~Mpc$^{-1}$).
   
\section{Sample Selection}

\subsection{Multi-wavelength Data}

We take as our initial GOODS-S sample all MIPS sources detected at
24~\micron\ to a flux density of $f_{24~\micron} > 80.0$~\microjy.  At
this flux limit, 99\% of the MIPS sample are detected to $> 10\sigma$.
There are several advantages to choosing a flux-limited MIPS
sample. First, AGN (and LIRGS/ULIRGS) tend to be bright at 24~\micron\
(e.g. Rigby et al. 2004).  Selecting only those galaxies with
$f_{24~\micron} > 80.0$~\microjy\ therefore retains all but the
faintest AGN while excluding 50-60\% of IRAC-selected IR-normal
star-forming galaxies at all flux densities.  Second, a MIPS
flux-limited sample is not subject to the complicated slope-dependent
selection bias present in IRAC selected samples due to the significant
variation in the sensitivity of the 4 IRAC bands (see Donley et
al. 2007). Third, this selection gives a complete and well-defined
sample of objects with extreme red $R - [24]$ colors.

The MIPS 24~\micron\ catalog of the GOODS Legacy team (Dickinson et
al., in prep.)  is comprised of 948 sources in the MIPS CDF-S Legacy
field with $f_{\nu} > 80.0$~\microjy. While this MIPS depth can be
obtained over the full CDF-S, we chose to limit this study to the
GOODS region to take advantage of the super-deep IRAC imaging.  The
relative depths of the limiting MIPS flux and the super-deep IRAC
photometry ensure that essentially all MIPS sources have high S/N IRAC
SEDs, allowing us to study in an unbiased way the IRAC properties of
this flux-limited MIPS sample.  To ensure that all AGN candidates have
sufficient X-ray coverage, we also required X-ray coverage of $T_{\rm
X} > 250$~ks from the deep $1~Ms$ CDF-S X-ray dataset (see Giacconi et
al. (2002) and Alexander et al. (2003)). Despite this relatively low
cut, 96\% of the final MIPS sample have $T_{\rm X} > 0.5$~Ms and 80\%
have $T_{\rm X} > 0.75$~Ms.  The resulting sample was drawn from an
area of 195.3 sq. arcmin and contains 846 MIPS sources.

The CDF-S is one of the best-imaged fields in the UV, optical, NIR,
and X-ray. We took advantage of this extensive multiwavelength dataset
by creating an aperture-matched catalog using the UV-NIR photometry of
Marzke et al. (1999, $RIz$), Vandame et al. (2001, $JK$), Arnouts et
al. (2002, $UU_{\rm p}BVRI$), COMBO17 (Wolf et al. 2004), Giavalisco
et al. (2004, $bvizJHK$), Le F{\`e}vre et al. (2004, $I$), and GALEX
($FUV,NUV$) (see P{\'e}rez-Gonz{\'a}lez et al. 2005 and the UCM
Extragalactic
Database\footnote{http://t-rex.fis.ucm.es/~pgperez/Proyectos/databaseuse.en.html}
for more details).  We then removed from our sample the 89 MIPS
sources that had multiple optical counterparts that were (1) within a
2.5\arcsec\ search radius of the MIPS source and (2) separated by more
than 0.5\arcsec.  (At $r < 0.5$\arcsec, it is difficult to distinguish
between multiple counterparts and an extended/irregular source with
multiple components.)  While we do not restrict our sample to regions
covered by GOODS ACS imaging, 83\% of the MIPS sources in our sample
have deep ACS coverage.

SExtractor-selected IRAC sources were similarly matched to the MIPS
sample after combining the super-deep IRAC data with data from the
deep, wide-area, \textit{Spitzer} Legacy Program (PI: van Dokkum) and
the MIPS-GTO IRAC program (see P{\'e}rez-Gonz{\'a}lez et al. 2008 for
further details).  To ensure accurate MIR SEDs, we removed from the IR
\plaga\ and color-selected samples 32 sources with blended
IRAC or MIPS photometry. Of the remaining 725 MIPS sources that meet
our criteria, 713 (98\%) have unique IRAC counterparts within a
2\arcsec\ search radius, and 699 (96\%) have $> 5\sigma$ IRAC
detections in all 4 IRAC bands, allowing us to determine accurately
the MIR colors of essentially all members of the MIPS-selected sample.
Of the 12 sources without IRAC counterparts, 6 were not detected due
to blending with a nearby source, 4 had badly centered MIPS positions,
and 2 had faint IRAC counterparts that fall below our catalog limit.

\subsection{Power-law, color-selected, and IR-normal samples}

After assigning IRAC counterparts, we separated the MIPS sample into
three subsets: IR \plagas, IR color-selected galaxies, and IR-normal
galaxies.  We defined as \plagas\ sources whose 4-band IRAC photometry
is well-fit by a line of slope $\alpha \le -0.5$, where $f_{\nu}
\propto \nu^{\alpha}$ (Alonso-Herrero et al. 2006; Donley et al. 2007).  
The effect of different cuts in $\alpha$ will be discussed in \S6.  To
ensure a good fit, we required the chi-squared probability P${\chi}$
(the probability that the fit would yield a chi-squared greater than
or equal to the observed chi-squared) to exceed 0.1.  P${\chi}$ tends
either to lie close to 0.5 (the probability that corresponds to a
reduced chi-squared of 1) or is very small (see Bevington \& Robinson
2003).  This selection, which identified 55 \plagas, was done using
the {\sc linfit} task in IDL.  This task takes the following 4-band
input from IRAC:

\begin{equation}
x=\rm{log}(\nu)
\end{equation}
\begin{equation}
y=\rm{log}(f_\nu)
\end{equation}
\begin{equation}
\Delta y=[1/\rm{ln}(10)]*\Delta f_{\nu}/f_{\nu}
\end{equation}

\noindent 
and returns the best fit slope, $\alpha$, and the chi-squared
probability, P${\chi}$.  While sources selected via this method are
dominated by the AGN in the mid-IR, we do not require that the
power-law extend into the optical.  Consequently, many
infrared-selected \plagas\ are dominated by stellar emission at
wavelengths short of $\sim 2$~\micron, where the reprocessed emission
from hot dust is suppressed because of dust sublimation.

Color-selected galaxies were defined as sources that meet the AGN IRAC
color-cuts of Lacy et al. (2004) or Stern et al. (2005), but that {\bf
do not meet the \plaga\ criterion}.  As discussed in Alonso-Herrero et
al. (2006) and Donley et al. (2007), both the \plaga\ and IRAC color
cuts attempt to select luminous AGN that outshine their host galaxies
in the infrared, filling in the dip in a galaxy's SED between the
short-wavelength stellar emission feature and the long-wavelength dust
emission features.  As such, the IRAC AGN color selection regions
contain, but also extend beyond, the power-law locus in color
space. While nearly all \plagas\ meet the IRAC AGN color selection
criteria, not all color-selected galaxies meet the \plaga\ criteria.
We therefore separate color-selected sources that can be identified
via a power-law fit (\plagas) from those that can not (color-selected
galaxies). The color-selected sample consists of 210 sources, 188 of
which meet the Lacy et al. criteria, 72 of which meet the Stern et
al. criteria, and 50 of which meet both criteria.

Finally, we define IR-normal galaxies as sources that meet neither the
IRAC \plaga\ nor the color selection criteria; these sources comprise
the remaining 448 galaxies in the MIPS sample.  As they are not the
focus of this study, we have not checked their IRAC and MIPS
photometry by eye. Instead, we estimate that the fraction of IR-normal
galaxies with blended IRAC or MIPS photometry is similar to that found
for the \plaga\ and color-selected samples, 11\%, or $\sim 50$
galaxies. We caution that IR-normal does \textit{not} mean purely
star-forming.  Instead, 'IR-normal' only indicates that any mid-IR
emission from an AGN is overwhelmed by emission from the host-galaxy.
In fact, many Type 2 and Seyfert-luminosity AGN meet the IR-normal
criteria (Stern et al. 2005, Donley et al. 2007, Cardamone et
al. 2008).

\subsection{IR-Excess Galaxies}

In addition to dividing the MIPS-selected sample into the 3
sub-samples discussed above, we identified IR-excess galaxies using
the criteria of Daddi et al. (2007a), Dey et al. (2008), Fiore et
al. (2008), and Polletta et al. (2008).  (We also searched for bright
ULIRGS that met the IR-excess criteria of Yan et al. (2007), but found
none in our faint sample.)  Daddi et al. (2007a) selected galaxies in
GOODS (with a $3 \sigma$ 24~\micron\ flux limit of 15-30~\microjy)
whose total MIR+UV star-formation rate (SFR) exceeds the
dust-corrected UV SFR by a factor of $>3$, and estimate that at least
$\sim 50\%$ of their sample are Compton-thick AGN.  Dey et al. (2008)
select sources with $R-[24] \ge 14$ ($f_{\rm 24~\micron}/f_{R} \gsim
1000$) and $f_{\rm 24~\micron} > 300$~\microjy, criteria which yield a
sample of both heavily obscured AGN and star-forming galaxies.  Fiore
et al. similarly require that $f_{\rm 24~\micron}/f_{R} \ge 1000$, but
also include an optical/NIR criterion of $R-K > 4.5$ and extend their
selection to fainter 24~\micron\ fluxes of $f_{\rm 24~\micron} \ge
40$~\microjy.  They estimate from simulations that 80\% of the sources
selected via these criteria are obscured AGN. Polletta et al. (2008)
focus only on the most luminous AGN ($f_{\rm 24~\micron} \gsim 1$~mJy)
with large infrared to optical flux ratios, whose IRAC and MIPS colors
can be described by the following criteria: $f_{\rm 5.8}/f_{\rm 3.6} >
2, f_{\rm 8.0}/f_{\rm 4.5} > 2$, and log $[f_{\rm 8.0}/f_{\rm 3.6}] +
\rm{log} [f_{\rm 24}/f_{\rm 3.6}] > 2$. 

Because IR-excess sources tend to be optically-faint, we use the
approach of Fiore et al. (2008, private communication) and estimate
the R-band magnitudes by interpolating the ACS v- and i-band data from
the MUSIC catalog (Grazian et al. 2006).  Despite the inherent
uncertainties associated with this relatively simple method, the
interpolated R-band magnitudes are in excellent agreement with
ground-based R-band measurements for bright sources, and they greatly
improve upon the uncertain R-band estimations at faint flux densities.
Because the MUSIC catalog is based on identifications at $z$ and $K$
(the latter of which is universally bright for the IR-excess sources),
the use of the MUSIC catalog for these sources also ensures that the
correct optical counterpart is chosen, as verified by a visual
inspection of the sources selected via the Fiore et al. criteria.  The
only disadvantage of this method is that, of the 195.3 sq. arcmin of
our survey, only 132.7 sq. arcmin (68\%) are covered by the deep ISAAC
K-band data, which was also used in the selection of the Daddi et
al. sources.  Our identification of the Dey et al., Daddi et al., and
Fiore et al. IR-excess samples is therefore limited to this region.

Of the 465/713 sources in our MIPS-selected sample that lie in the
ISAAC field, 10 meet the Dey et al. criteria, 52 meet the Fiore et
al. criteria, and 42 lie in the Daddi et al. IR-excess sample (the
list of which was kindly provided by D. Alexander, private
communication 2008). In addition, 71 MIPS sources have red IR/optical
colors of $f_{\rm 24~\micron}/f_{R} \ge 1000$; we will refer to this
sample as 'IR-bright/optically-faint'.  Of the full sample of 713
sources, 5 meet the Polletta et al. criteria.  The properties of these
IR-excess galaxies, nearly all of which also meet the power-law or
color-selection criteria outlined above, will be discussed in more
detail in \S7.

\section{Redshifts}

While the redshift coverage in the CDF-S is amongst the highest in all
cosmological fields, only 34\% of the sources in our faint sample have
spectroscopic redshifts.  As one of our main goals is to investigate
the redshift-dependency of infrared color selection, we require both
accurate and complete redshift information. However, the mean
magnitude of the sources without spectroscopic redshifts, $V\sim
24.1$, is very faint, making further spectroscopic follow-up
challenging. We therefore supplement the spectroscopic redshifts with
photometric ones. The sources that are the focus of this study
generally have SEDs with weak stellar features and require extra care
to fit accurate photometric redshifts.  To insure this accuracy, we
have used two complementary photo-z approaches and have incorporated
two stages of independent visual inspection by two reviewers, as
described in Appendix A.  The final results, shown in Figure 1,
support the high accuracy and completeness of the adopted redshifts
for this faint sample.

\begin{figure}
\epsscale{1}
\plotone{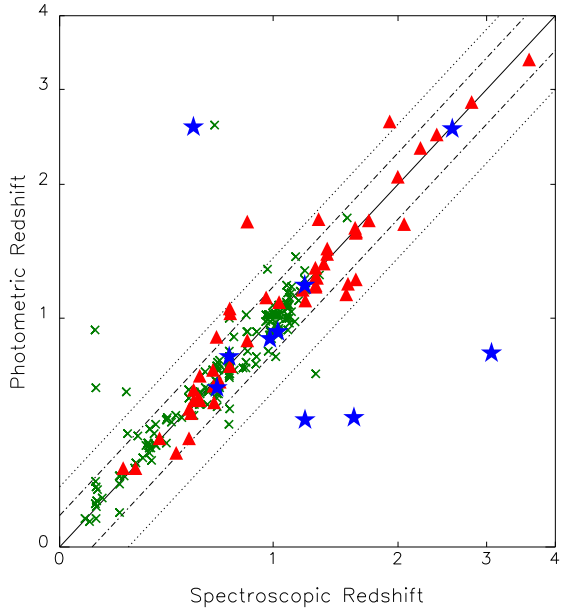}
\caption{Photometric vs. secure spectroscopic redshifts for
\plagas\ (navy stars), color-selected galaxies (red
triangles), and IR-normal galaxies (green crosses). Values are plotted
in log$(1+z)$. Dot-dashed and dotted lines give 10\% and 20\% errors
on $\Delta z$, respectively. 89\% of the sample have
$\Delta(z)$$<$0.1, and 95\% have $\Delta(z)$$<$0.2.}
\end{figure}

Of the 713 MIPS-selected sources in our sample, 249 have secure
spectroscopic redshifts from the VVDS (Le F{\`e}vre et
al. 2004),VLT/FORS2 (Vanzella et al. 2006), K20 (Mignoli et al. 2005),
and Szokoly et al. (2004) redshift surveys.  For our photometric
redshifts, the mean offset is $\overline{\Delta (z)} = 0.012$, where
$\Delta(z) = (z_p - z_s) / (1+z_s)$, and the dispersion is
$\sigma_{\rm z} = 0.15$, where $\sigma_{\rm z}^2 = (1/N) \sum \Delta
(z)^2$. Eighty-nine percent of the sample have $\Delta(z)$$<$0.1, and
95\% have $\Delta(z)$$<$0.2.  As shown in Table 1, while the redshift
completeness and accuracy are high for the IR-normal and
color-selected samples, they drop significantly for the \plaga\
sample.  A comparison of their spectroscopic and photometric redshifts
illustrates some of the issues. Three of the four \plaga\ outliers are
best fit by a Type 1 QSO template, whose redshift is particularly
difficult to constrain. However, of the 18 \plagas\ for which only
photometric redshifts are available, only 2 (11\%) are fit by a Type 1
QSO template. The remainder show optical features/breaks that make
their redshift determination more secure. As such, we expect the
overall accuracy of the \plaga\ redshifts to be higher than one would
assume given the limited comparison with spectroscopic redshifts.

With the addition of 400 photometric redshifts, computed as described
above, our redshift completeness is 91\%. The redshift distribution is
shown in Figure 2. Also plotted in Figure 2 is a redshift histogram
that incorporates typical errors in the photometric redshifts.  To
produce this distribution, we simulated 10,000 redshift distributions
in which the photometric redshifts were randomly varied according to
the appropriate $\sigma$ given in Table 1. We plot in Figure 2 the
mean of the resulting distributions.  Both redshift distributions show
a strong peak at $z=1$, and while the first shows a potential peak at
$z\sim 2$, we cannot confirm its presence due to the errors on the
photometric redshifts.

\begin{figure*}
\epsscale{1}
\plottwo{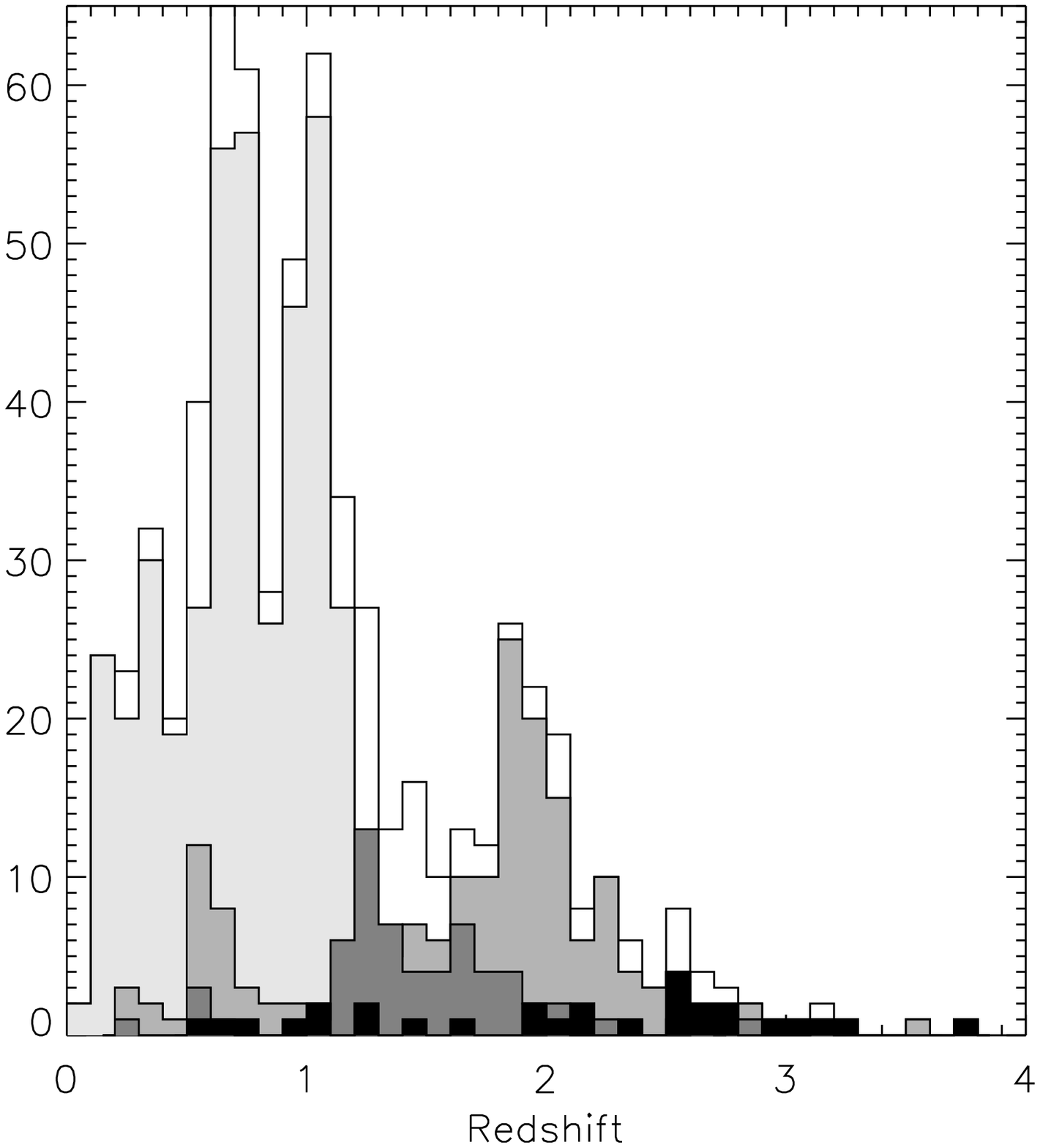}{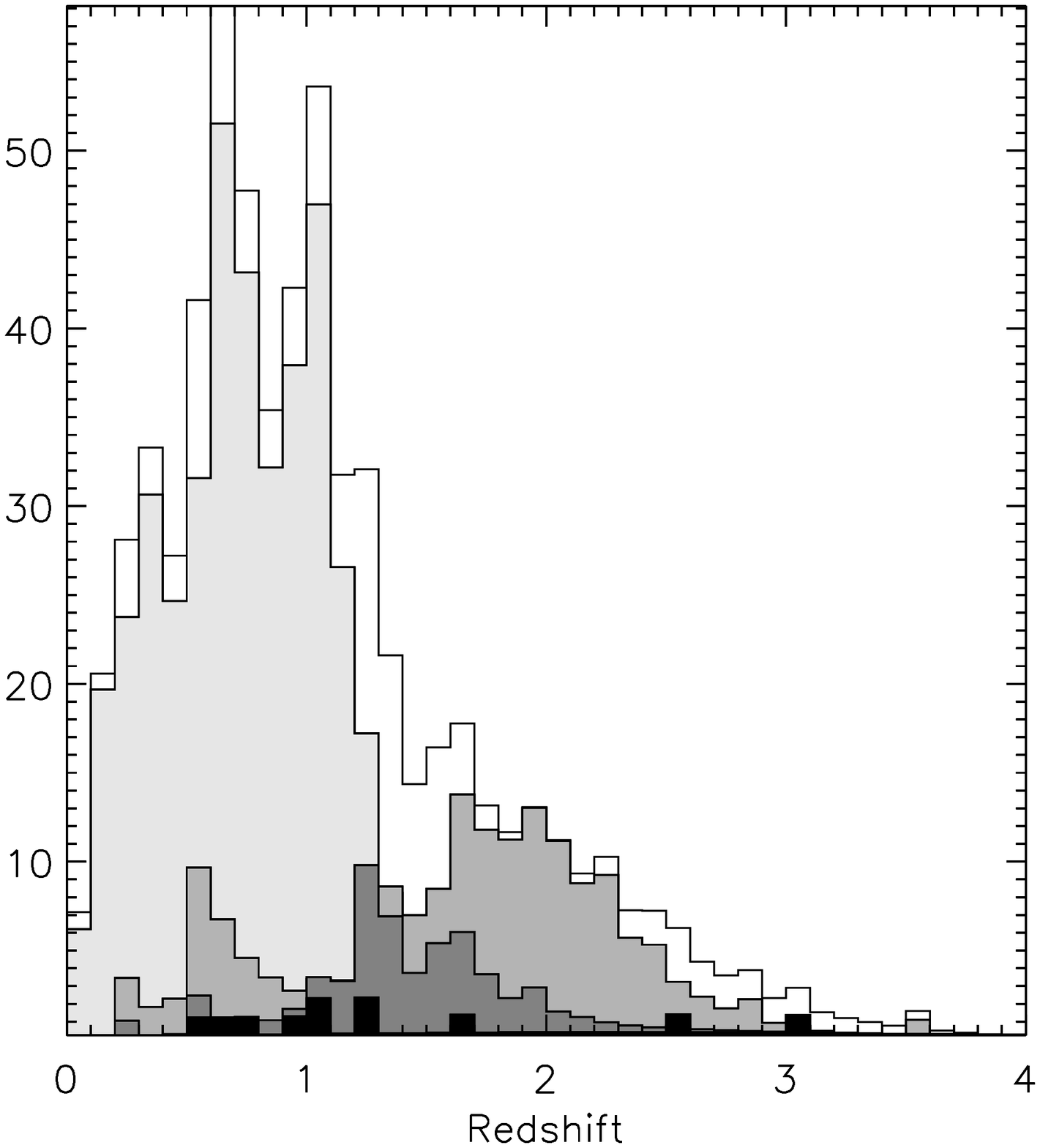}
\caption{Redshift distribution of the MIPS-selected sample.  The plot on the left
gives the observed distribution, while the plot on the right
incorporates the typical errors on the photometric redshifts. From
lightest to darkest shading, the histograms represent all MIPS sources
in our sample, IR-normal galaxies, color-selected sources that meet
the Lacy et al. (2004) criteria, color-selected sources that meet the
Stern et al. (2005) criteria, and
\plagas.}
\end{figure*}

We separate the sample in Figure 2 into the \plaga, color-selected
(Lacy et al. and Stern et al.), and IR-normal subsamples.  As
expected, the number of IR-normal galaxies peaks at $z=0.7$ (a
well-known redshift peak in the CDF-S) and at $z=1.1$ (another known
redshift peak).  We also detect a peak at $z=0.3$ similar to that
found by Desai et al. (2008), whose strength increases if we apply
their cut of $f_{\rm 24~\micron} \ge 300$~\microjy.  In the CDF-S,
however, this peak is dwarfed by the stronger peak at $z=0.7$, even at
large flux densities.  The redshift distribution of the IR-normal
galaxies decreases rapidly at redshifts of $z > 1.2$, where only
highly luminous star-forming galaxies are detectable. In contrast, the
\plagas\ with redshift estimates have a relatively flat distribution
in redshift space, as was found for the \plagas\ in the CDF-N (Donley
et al. 2007).

The redshift distributions of the (non-power-law) Lacy et al.  and
Stern et al. selected samples differ significantly, both from the
IR-normal and \plaga\ samples, as well as from one another.  The
Stern-selected galaxies peak at $z \sim 1.25$ whereas the
Lacy-selected galaxies show significant peaks at $z \sim 0.5$ and $z
\sim 2$, with very few galaxies falling in the $z<1$ regime originally 
probed by these selection methods (see \S5).  In addition, at $z \ge
1.75$, nearly all (94\%) MIPS-selected sources meet the Lacy AGN
selection criteria, regardless of their nature.  A large concentration
of galaxies at $z \sim 2$ was previously observed in this field by
Caputi et al. (2006), and in a brighter sample of MIPS sources by
Desai et al. (2008), and is most probably due to the 7.7~\micron\
aromatic feature passing through the 24~\micron\ band.  This behavior
suggests that the mid-IR continua of the Lacy color-selected galaxies
contain substantial contributions from star-formation (e.g. Genzel et
al. 1998).

Finally, the mean redshifts of the IR-excess samples are as follows:
$z=1.92 \pm 0.36$ for the Daddi et al. sources, $z=2.19 \pm 0.61$ for
the dust obscured galaxies (DOGs, Dey et al. 2008), $z=2.09 \pm 0.48$
for the Fiore-selected sources, $z=2.11$ for the one Polletta source
with a known redshift, and $z=2.05 \pm 0.49$ for the
IR-bright/optically-faint sources. As discussed above, this
concentration about a redshift of 2 (when not a design of the
selection as in Daddi et al. (2007)) is likely to be due at least in
part to the passage of the 7.7~\micron\ aromatic feature through the
MIPS 24~\micron\ band, suggesting a significant contribution from
star-formation.

\section{X-ray Properties} 

Of the 713 MIPS sources, 109 (15\%) have X-ray counterparts in the
Alexander et al. (2003) or Giacconi et al. (2002) catalogs.  Of these,
25 are \plagas, 35 are color-selected galaxies (33 from the Lacy
criteria, and 12 from the Stern criteria), and 49 are IR-normal.
While the IR-normal galaxies therefore dominate the X-ray counts, the
fraction of such sources is low: only 11\% of the IR-normal galaxies
have X-ray counterparts, as compared to 17\% of the color-selected
galaxies and 45\% of the \plagas.

We test for faint ($\ge 2\sigma$) X-ray emission from the MIPS sources
using the procedure outlined in Donley et al. (2005).  The resulting
detection fractions for the \plaga, color-selected, IR-normal, and
IR-excess samples are given in Table 2.  With the inclusion of the
weakly-detected X-ray sources, the detection fractions of the
IR-normal, color-selected, and \plaga\ samples increase to 40\%, 42\%,
and 64\%, respectively.  As indicated in Table 2, however, while all
strongly and weakly-detected \plagas\ have AGN X-ray luminosities of
log~$L_{\rm x}$(ergs~s$^{-1}$)$ \ge 42$, the same is true for only
74\% of the color-selected galaxies, and 34\% of the IR-normal
galaxies.  Given the infrared luminosities implied by our 24~\micron\
selection criterion, the portion of the sample with log~$L_{\rm
x}$(ergs~s$^{-1}$)$ < 42$ will be heavily contaminated with
star-forming galaxies (Ranalli et al. 2003).

While the X-ray detection fraction of \plagas\ is relatively low
(though significantly higher than that of the color-selected
galaxies), it is comparable to that of previously-selected samples
(Alonso-Herrero et al. 2006, Donley et al. 2007).  Of the \plagas\ in
the 2~Ms CDF-N, 55\% had high-significance X-ray counterparts.
However, only 15\% remained undetected down to the 2.5$\sigma$
detection level, suggesting that many \plagas\ are likely to be
heavily obscured X-ray sources whose fluxes fall below the current
detection limits (Donley et al. 2007).  In the CDF-S, only 45\% of the
\plagas\ have cataloged X-ray counterparts and 36\% remain undetected
down to $2\sigma$.  This slightly lower detection fraction is due at
least in part to the lower X-ray exposure of the 1~Ms CDF-S, and will
be discussed further in \S6.

The X-ray detection fraction of the luminous IR-excess sources
selected via the Polletta et al. (2008) sample is high (80\%).
However, the same can not be said for the remaining IR-excess samples:
only 30\% of the DOGs, 19\% of the Fiore et al. sources, and 15\% of
the sources with high 24~\micron\ to optical flux ratios have
cataloged X-ray counterparts. By definition, none of the Daddi et
al. sources have cataloged hard X-ray counterparts, although 3 have
soft-band counterparts. When the weakly-detected X-ray counterparts
are included, these numbers rise to 100\% for the Polletta sources,
63\% for the DOGs, 44\% for the Daddi et al. sources, 43\% for the
Fiore et al. sources, and 42\% for the IR-bright/optically-faint
sources.

\section{IRAC color-color selection}

There are several reasons why a re-examination of the color-color
selection criteria is needed before applying these techniques to our
sample.  First, thanks to the availability of high quality NIR and
\textit{Spitzer} MIR spectra and photometry, we can now construct more
accurate MIR templates for both AGN and star-forming galaxies than
were available when the Lacy et al. and Stern et al. AGN selection
criteria were initially defined.  Second, the AGN selection criteria
of Lacy and Stern were initially designed for use with shallow
surveys, and have not yet been properly tested over a range of both
redshift and flux density (but see Cardamone et al. 2008).  Third,
photometric redshift techniques for \textit{Spitzer}-detected galaxies
have advanced sufficiently to allow nearly complete redshift
estimation.

The positions of the MIPS-selected sample in IRAC color space are
shown in Figure 3, where stars represent the \plagas\ and circles
represent the remaining MIPS sources. As expected, all
\plagas\ meet the Lacy criteria and all but 3 meet the Stern
criteria.  The relatively large scatter of the \plagas\ on the Stern
plot probably arises in part from the use of adjacent color bands,
whereas the power-law fitting tends to smooth over noise.

\begin{figure*}
\plottwo{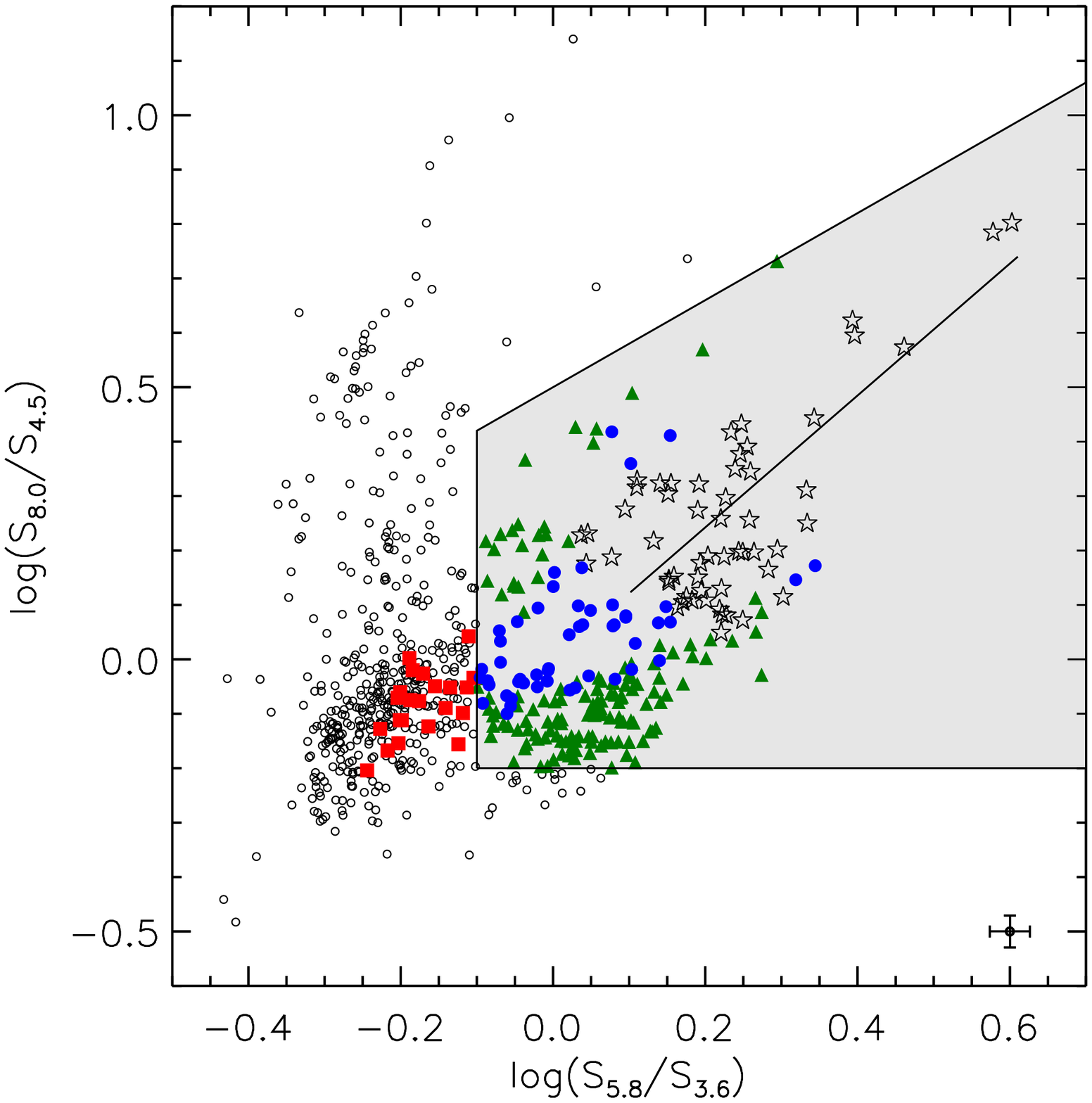}{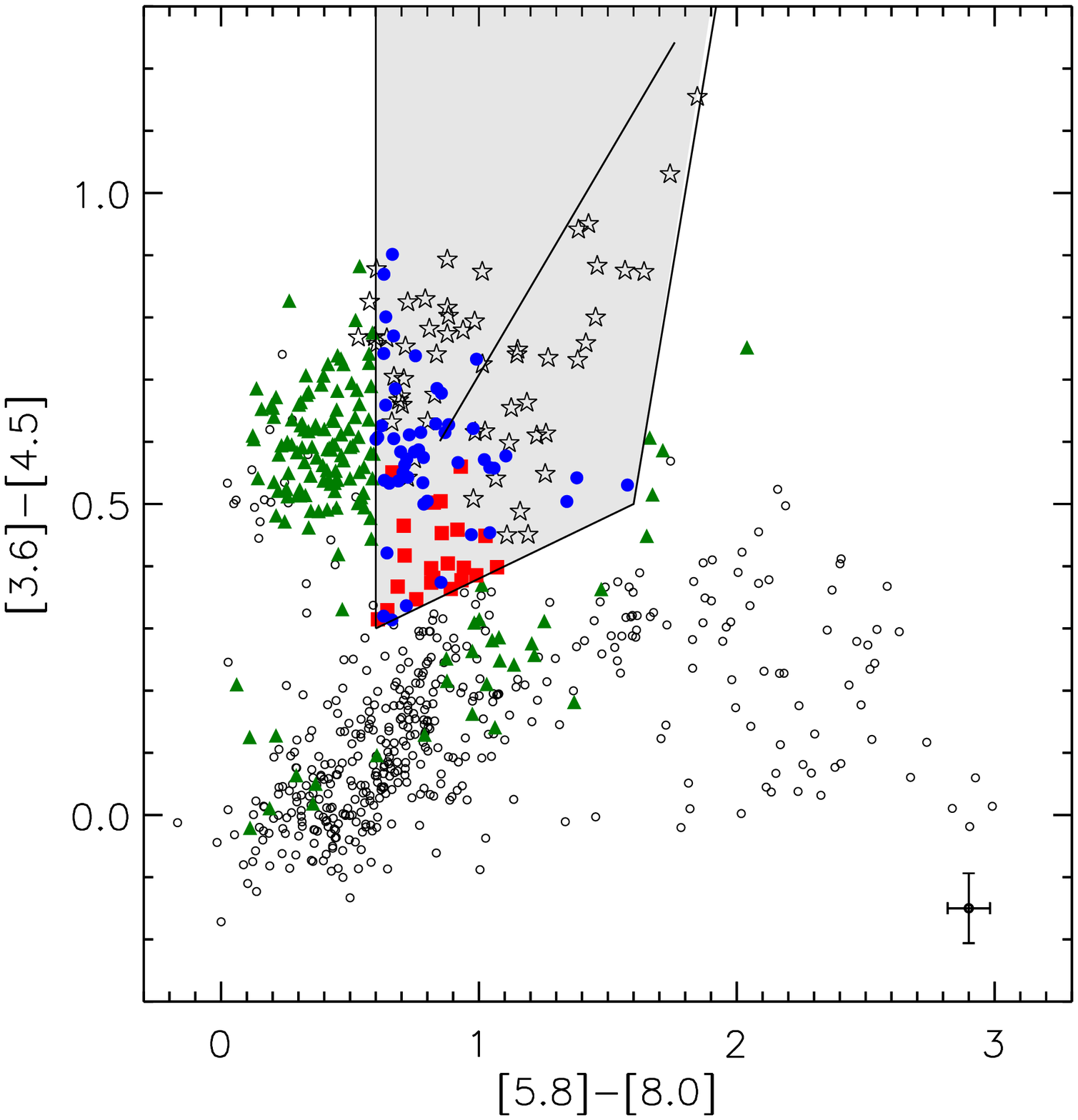}
\caption{Position in Lacy et al. (2004; left) and Stern et al. (2005; right) 
color-space of the MIPS-selected sample, where \plagas\ are given as
stars. Blue filled circles represent color-selected galaxies that meet
both the Lacy et al. and Stern et al. criteria.  Green triangles and
red squares represent sources that meet only the Lacy et al. or Stern
et al. criteria, respectively.  The small open circles are IR-normal
galaxies. The shaded regions represent the AGN selection regions, and
the diagonal lines within are the loci of perfect power laws with
$\alpha = -0.5$ to $-3.0$.}
\end{figure*}

\vspace*{1cm}
\subsection{Comparison of the IRAC color-color selection criteria}

To illuminate the strengths and weaknesses of the two color-color
selection criteria in Figure 3, we separate the sources that meet both
the Lacy et al. (2004) and Stern et al. (2005) AGN selection criteria
from those that meet only one of the two criteria.  The color-selected
sources that meet both criteria primarily define an extension of the
power-law locus to a blue slope of $\alpha = +0.5$ (recall that our
definition of \plagas\ includes only those sources with red slopes of
$\alpha < -0.5$).  We will discuss in detail the effects of different
power-law slope criteria in
\S6.

The Lacy-only sources occupy two regions in color-space.  The first,
located in the lower portion of the Lacy diagram, corresponds to the
concentration in the upper-left corner of the Stern et al. selection
region.  These sources were intentionally excluded from the Stern
selection region to minimize contamination from high-redshift ($z\sim
2$) star-forming galaxies.  The second concentration of Lacy-only
galaxies is scattered throughout the star-forming locus of the Stern
diagram.  Are these normal, low-z star-forming galaxies, or obscured
AGN not selected by the Stern criteria?

The Stern-only sources fall almost exclusively in the lower left
portion of the Stern wedge, and occupy the region dominated by
low-redshift IR-normal galaxies in the Lacy et al. color diagram,
suggesting that, like the Lacy criteria, the Stern et al. criteria are
likely to suffer from low-redshift star-forming galaxy contamination.
Because of the exclusion of $z \sim 2$ star-forming galaxies, however,
the Stern et al. criteria should perform better at high $z$.  To test
this hypothesis, we next consider the evolution in IRAC color-space of
a number of high-quality star-forming and AGN templates.

\vspace*{0.1cm}
\subsection{Star-forming Templates}

As discussed above, the availability of high-quality near- and
mid-infrared data has allowed the construction of high-quality
star-forming SEDs, particularly for luminous and ultra-luminous
infrared galaxies (LIRGS and ULIRGS).  In Figure 4, we plot the
redshift evolution of these templates in IRAC color-color space over
$z=0-4$, where purple and blue tracks represent the \textit{purely
star-forming} ULIRG and LIRG templates of Rieke et al. (2008, see
Table A1), green tracks represent the spiral and starburst templates of
Polletta et al. (2007) and Dale \& Helou (2002), and red tracks
represent the elliptical templates of Silva et al. (1998).  Large
circles mark the tracks at $z=0$, and small circles mark each integer
redshift from $z = 1$ to 4.

\begin{figure*}
\plottwo{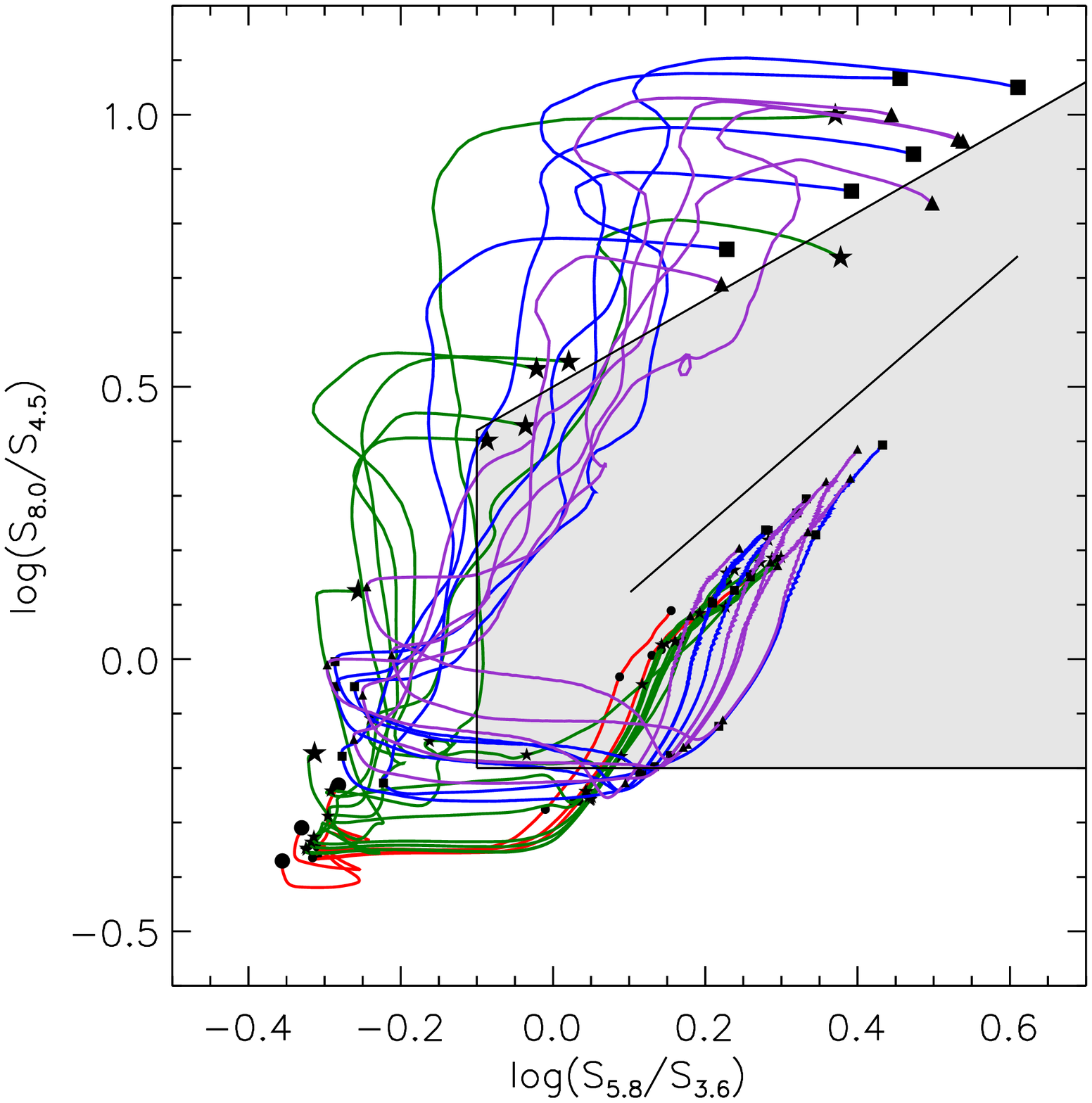}{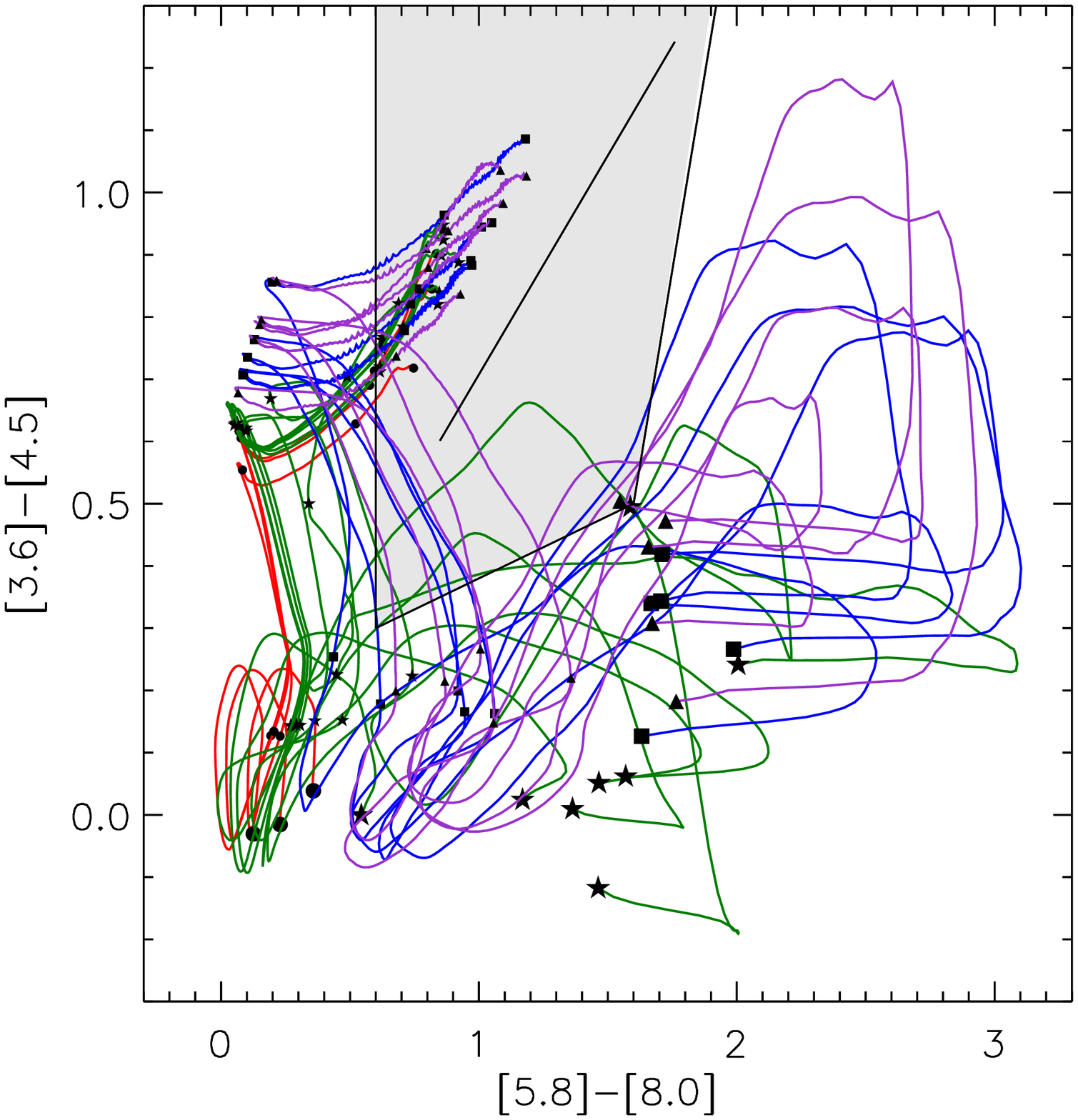}
\caption{Tracks in color-space of the \textit{purely star-forming} SEDs 
of ULIRGS (purple, triangles), LIRGS (blue, squares), spirals and
starbursts (green, stars), and elliptical galaxies (red, circles),
from redshifts of $z=0$ (large symbols) to $z=4$. Small symbols mark
redshift intervals of 1.  The power-law locus with $\alpha = -0.5$ to
$-3.0$ is shown as a line inside the shaded AGN selection regions. The
star-forming SEDs enter the AGN selection regions at both low and high
redshift.}
\end{figure*}

While the star-forming templates generally avoid the power-law locus
itself, they enter the Lacy and Stern selection regions at both low
and high redshifts, tracing out the same regions in color space
occupied by many of the color-selected AGN, particularly those
selected via only one of the two criteria.  The templates therefore
suggest potential star-forming galaxy contamination of the
color-selected AGN, as previously predicted by Barmby et al. (2006),
Donley et al. (2007), and Cardamone et al. (2008), and indicate that
the current AGN selection regions may inadequately separate AGN and
star-forming galaxies.

Our results can be compared with the simulations of Sajina, Lacy, \&
Scott (2005), who calculated mathematical models of galaxies from
three spectral components: stars, aromatic features, and a
continuum. For $0 < z < 1$, we agree with their Figure 8 (upper left)
that starburst luminosity galaxies do not significantly 'invade' the
Lacy AGN color wedge (see our Figure 4). However, we find that more
luminous galaxies can invade this wedge much more seriously (again see
Figure 4). The difference may arise because the inputs to their models
included few star-forming LIRGs and only one ULIRG (Arp 220, an
atypical case). Therefore, it is likely that the behavior of the most
luminous star forming galaxies is not captured as accurately in their
models as is that of lower luminosity ones. By $z = 1$, the typical
\textit{Spitzer} 24~\micron\ survey sensitivity limit reaches only to the
bottom of the LIRG range, so the Lacy AGN color wedge is more
susceptible to contamination than was concluded by Sajina et
al. (2005).

\subsection{Redshift-dependent color selection}

While the star-forming templates appear to trace quite well the
positions of many of the color-selected galaxies in color-color space,
Figure 4 covers a wide range of redshifts ($z=0-4$).  To understand
better the overlap between the star-forming templates and the
color-selected galaxies (and \plagas), we must take the redshift
information into account. We therefore break the sample down into
smaller redshift bins, as shown in Figures 5 and 6. We overplot on the
color-color diagrams the redshift-appropriate colors of purely
star-forming galaxies.  To simplify the plots, we do not plot each
galaxy track separately, as was done in Figure 3, but instead draw 1
$\sigma$ contours around the tracks, where $\sigma$ is taken to be the
median IRAC measurement error of the full MIPS sample. We find that,
unlike the \plagas, the majority of the color-selected AGN candidates
fall within or very close to the contours for star-forming galaxies of
similar redshifts.  Thus, it is likely that their mid-IR SEDs are
dominated by star formation.  This result suggests that simple mid-IR
color-color cuts cannot identify reliable AGN samples without also
including redshift-based or additional SED (e.g. power-law) criteria.
We discuss in Appendix B the individual redshift intervals.

\begin{figure*}
\plotone{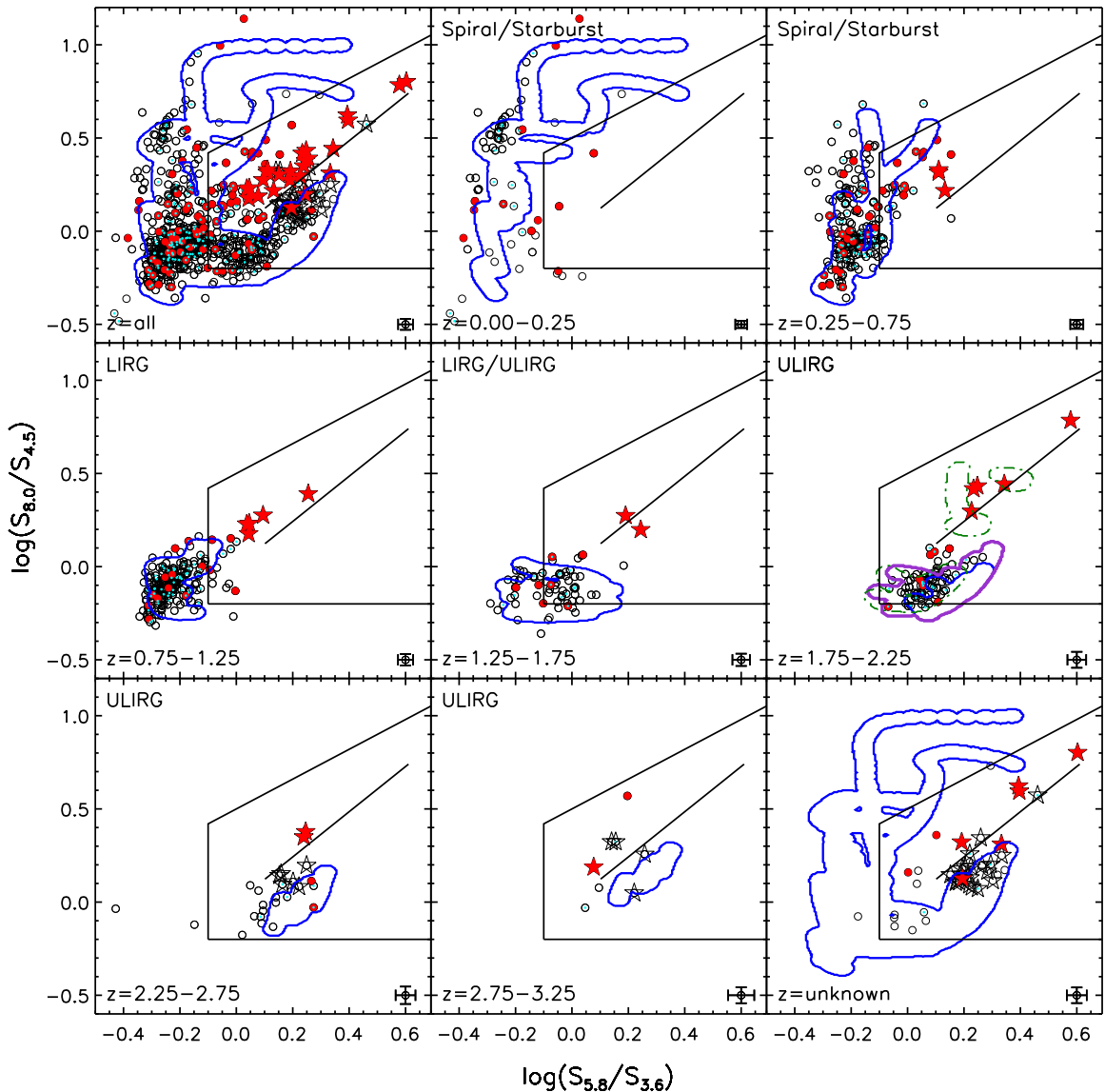}
\caption{Position in Lacy et al. (2004) color-space of the MIPS-selected sample, 
as a function of redshift, where \plagas\ are shown as stars,
X-ray--cataloged sources are given by filled (red) symbols, and X-ray
weakly-detected sources are given as small cyan symbols.  Overplotted
are the redshift-appropriate contours representing the IRAC colors of
purely star-forming templates, assuming no errors on the photometric
redshifts.  The thick (purple) contours and dot-dashed (green)
contours in the $z=1.75-2.25$ redshift bin represent star-forming
galaxy templates for which 10\% errors were incorporated into the
redshift range, and AGN templates, respectively.}
\end{figure*}

\begin{figure*}
\plotone{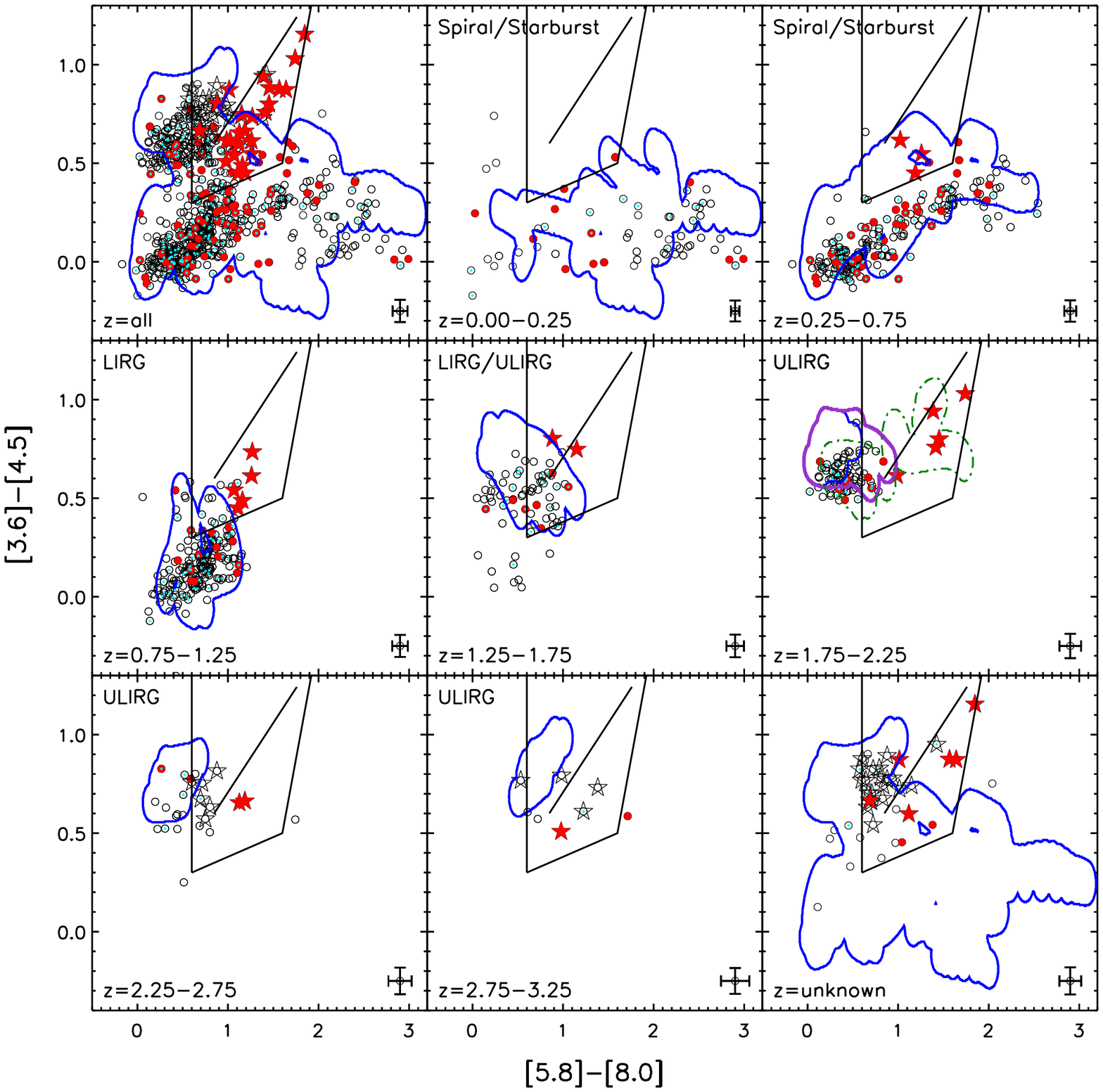}
\caption{Position in Stern et al. (2005) color-space of the MIPS-selected sample, 
as a function of redshift.  Symbols and contours are as described in
Figure 5.}
\end{figure*}

The summary of our findings can be found in Table 3, where we present
the overall fraction of color selected galaxies and \plagas\ that lie
outside the 1, 2, and 3$\sigma$ star-forming contours.  The top half
of the table gives the fractions assuming no errors on the photometric
redshifts; the lower half incorporates 10\% errors which for clarity
are imposed on the templates rather than on the individual galaxy
measurements. For the \plagas, we give two percentages: the fraction
of sources that lie outside the star-forming contours in Lacy and
Stern color-space, respectively.

As is clear from Table 3, the fraction of color-selected galaxies that
lie outside the star-forming contours is lower than that of the
\plagas\ by a factor of 2-10 at all three levels of significance.  For
instance, while 57\% - 71\% of \plagas\ lie more than $3 \sigma$ from
the star-forming contours in Lacy color space, the same can be said
for only 5\%-10\% of the color-selected galaxies.  In addition,
increasing the significance from 1$\sigma$ to 3$\sigma$ has a far
greater effect on the reliability of the color-selected galaxies than
on that of the \plagas, especially in Lacy color-space, indicating
that the color-selected galaxies lie noticeably closer to the
star-forming templates than do the \plagas, as expected.  We note that
this analysis depends on the templates being used, as the addition or
removal of any one star-forming template will result in changes in the
numerical results presented above and in the table.  The overall
trends discussed above, however, will remain the same.

\subsection{Properties of Color-selected AGN candidates}

While the majority of color-selected AGN candidates lie inside the
star-forming contours (see Table 3), a number of non-power-law
color-selected galaxies meet the AGN selection criteria and have
colors inconsistent with those of our star-forming templates. Are
these AGN, as predicted?  How do their redshift distributions, X-ray
detection fractions, and numbers compare to those of the \plagas?  In
the following discussion, we define as 'secure' color-selected
galaxies (and \plagas) those AGN candidates that lie $>1\sigma$ away
from the redshift-appropriate star-forming templates.

As discussed above, and as shown in Figure 2, the redshift
distributions of the color-selected sources show a strong peak at $z =
2.0$, attributed to the 7.7~\micron\ aromatic feature, a
star-formation indicator.  In contrast, the \plagas\ have a relatively
flat redshift distribution, indicating little or no contribution from
aromatic features.  If we do not account for errors in the photometric
redshifts, the resulting sample of 'secure' color-selected galaxies
retains the large $z=2$ peak, suggesting a large star-formation
contribution.  If we incorporate 10\% errors on the redshifts of the
star-forming templates when selecting our secure candidates, however,
the redshift distributions of the remaining color-selected galaxies
become relatively flat, with mean redshifts of $z=1.59 \pm 0.83$ and
$z=1.62 \pm 0.84$ for the Lacy and Stern-selected sources,
respectively, suggesting that a significant fraction of the
high-redshift star-forming contaminants have been removed.  For
comparison, the mean redshifts and rms ranges of the secure \plagas\
selected in Lacy and Stern color-space are $1.98 \pm 0.87$ and $2.09
\pm 0.78$.

While the redshift distribution therefore suggests that many of the
star-forming contaminants have been rejected from the secure sample,
the X-ray detection fractions, shown in Table 4, suggest otherwise.
At distances of 1 and 2 $\sigma$ from the star-forming templates, the
X-ray detection fraction of the \plagas\ exceeds that of the
color-selected galaxies by factors of $>2-3$.
This factor, however, should be taken with reservations; X-ray sources
are far more likely to have spectroscopic redshift estimates, and are
therefore far more likely to be included in our secure sample, which
requires redshift information.  While this has a minimal effect on the
color-selected galaxies, for which the redshift completeness is high,
it has a large effect on the \plagas, boosting their probable X-ray
detection fraction.

A better comparison is therefore with the IR-normal galaxy population,
11\% of which are detected in the X-ray.  The X-ray detection
fractions of the secure color-selected AGN candidates (14-29\%),
exceed this value, but only by a factor of $\sim 1-3$.  In comparison,
the full sample of \plagas\ has an X-ray detection fraction of 45\%,
which exceeds that of the IR-normal galaxies by a factor of 4. Only at
the highest significance, $3 \sigma$ from the star-forming contours,
does the X-ray detection fraction of the secure color-selected
galaxies approach that of the full \plaga\ sample.

This significant offset in X-ray detection fraction could be due
either to a difference in intrinsic luminosity or to lingering
contamination of the color-selected sample by star-forming galaxies.
While the average X-ray luminosity of the secure \plagas, log~$L_{\rm
x}$(ergs~s$^{-1}$)$ = 44.1$, exceeds that of the secure Lacy and
Stern-selected sources detected in the X-ray, log~$L_{\rm
x}$(ergs~s$^{-1}$)$ = 43.8$ and 43.9, respectively, the offset is
relatively small, suggesting that the large discrepancy in the X-ray
detection fractions is not driven primarily by a systematic offset in
X-ray luminosity of the relevant AGN.  Instead, it likely arises from
the inclusion of star-forming galaxies, even in the $1\sigma$ 'secure'
color-selected population, suggesting that a larger cut in $\sigma$
(e.g. $3 \sigma$) is required to define a reasonably secure sample.

Also shown in Table 4 are the total number of X-ray selected galaxies
in both the secure color-selected and \plaga\ samples.  After applying
completeness corrections for the fraction of Lacy, Stern, and
\plaga\ X-ray sources that have redshifts and that can therefore be included in the 
secure sample (94\%, 83\%, and 76\%, respectively), the number of
secure X-ray selected \plagas\ exceeds that of the secure
X-ray--detected color-selected galaxies, regardless of whether we
define the secure sample as those sources that lie 1, 2, or 3$\sigma$
from the star-forming contours.  This indicates that \textit{the
\plaga\ selection criterion identifies the overwhelming majority of
secure AGN candidates in IRAC color-space.}

\subsection{Flux dependency of color selection}

While the Lacy and Stern AGN selection criteria were defined using
relatively shallow surveys, these selection techniques are now being
applied to samples with a range of flux densities (e.g. Cardamone et
al. 2008).  How does the limiting flux affect the completeness and
reliability of AGN color selection?  Are the problems discussed above
present in shallow as well as deep surveys?

There are several reasons why we might expect to see a shift in
reliability with flux.  First, the AGN fraction of MIPS sources
depends quite strongly on the 24~\micron\ flux density (e.g. Treister
et al. 2006, Brand et al. 2006).  For instance, while only $\sim 4\%$
of MIPS sources at our flux limit of 80~\microjy\ are expected to be
AGN (Treister et al. 2006), the fraction at 5~mJy is more than an
order of magnitude greater ($\sim 45\%$).  Therefore, MIPS-selected
shallow surveys should contain fewer star-forming galaxies, although
those that remain are likely to be LIRGs/ULIRGs that enter the AGN
selection region in greater numbers than spirals and starbursts (see
Figure 4).

Second, many of the 'problem' sources that lie inside the AGN color
selection region but that also fall inside the star-forming contours
have moderately high-redshifts ($z>1.25$), and are therefore likely to
drop out of shallow surveys.  While shallow samples may therefore
contain high-luminosity ULIRGS, the exclusion of lower-luminosity
high-redshift galaxies reduces the risk of contamination.  Other flux
cuts, such as the $R$ magnitude $<21.5$ requirement of Stern et
al. (2005), also prevent contamination at high-$z$, as normal galaxies
at this magnitude are detected only to $z\sim0.6$.

To investigate the effect of intermediate flux density cuts on the AGN
color selection, we show in Table 5 the fraction of 'secure'
color-selected and \plagas\ (those that lie outside the 1$\sigma$
star-forming contours) as a function of flux density.  If we do not
incorporate errors in the photometric redshifts into our definition of
the secure sources, we find that the fraction of secure sources is
relatively constant at $\sim 50\%$ regardless of flux density.  If we
allow errors on the photometric redshifts when constructing our
'secure' sample, the reliability of Lacy-selected sources at $f\ge
80$~\microjy\ drops to 29\%, and that of Stern-selected sources drops
to 21\%.  At $f\ge 500$~\microjy, the fraction of secure sources
amongst both samples rises, but only to 50\%.  Regardless of our
assumptions, therefore, the fraction of potential contaminants is
still high ($\sim 50\%$), even at the highest fluxes probed by our
survey. Because of the pencil-beam nature of the CDF-S survey, 95\% of
the sources in our sample have $f_{\rm 24~\micron} \le 600$~\microjy,
so our ability to comment on brighter samples is limited.

It is also worth noting that the brightest MIPS sources tend to lie
above the power-law locus in Lacy color-space and to the right of the
power-law locus in Stern space, regions where we expect minimal
contamination from star-forming galaxies.  Not surprisingly, these
sources are also almost always detected in the X-ray, suggesting that
these regions of IRAC color-space are the most secure.

\subsection{Comparison with previous work}

In a study of 77 AGN candidates selected from the \textit{Spitzer}
First-Look (Lacy et al. 2005) and SWIRE surveys (Lonsdale et al. 2003)
via the Lacy et al. (2004) criteria, Lacy et al. (2007) found that
33\% are unobscured type 1 quasars, 44\% are type 2 AGN, and 14\% are
dust-reddened type 1 quasars. Only 9\% have star-forming or LINER
spectra.  Is this relatively low contamination by star-forming
galaxies consistent with our findings?

There are three main factors that lead to the high reliability of the
Lacy et al. (2007) sample.  First, the sample members are very bright,
with a typical 24~\micron\ flux density of 5~mJy.  As discussed above,
Treister et al. (2006) predict that 45\% of sources at this flux
density should be AGN, regardless of their MIR SEDs, compared to only
4\% of sources at our flux limit of 80~\microjy.  Indeed, of the
brightest 50 sources in the XFLS (whose median 24~\micron\ flux
density is 9.4 mJy), 58\% are optically classified as AGN (Lacy et
al. 2007). Thus, the sample from which these color-selected AGN was
drawn contains far fewer star-forming galaxies than the deeper GOODS
sample.

Second, while the sources in the Lacy et al. (2007) sample have
redshifts ranging from $z=0.053$ to 4.27, the median redshift of the
sample is low: $z=0.6$.  Only 5 color-selected sources have redshifts
in excess of $z=1.75$, the redshift above which nearly all sources in
our sample (AGN or star-forming) meet the Lacy et al. criteria.  We
would therefore expect little or no contamination by
\textit{high-redshift} star-forming galaxies, the predominant source
of contamination in our faint sample.

The third and most important reason for the high reliability of this
sample, however, is the high \plaga\ fraction of the Lacy-selected AGN
candidates.  Of the 77 sources in the Lacy et al. (2007) sample, 59
are \plagas\ as defined in \S2.  The 18 sources that are excluded by
the \plaga\ criteria include the only two starburst galaxies detected
in this subsample, one starburst/LINER, an unclassified high-redshift
galaxy with narrow UV emission lines, two composite galaxies, 8 Type 2
AGN (the identification of 4 of which were based on a BPT analysis), 3
reddened Type 1 AGN, and 1 Type 1 AGN.  If we relax the probability
constraint of the \plaga\ criterion to P${\chi} > 0.01$, as was done
in Alonso-Herrero et al. (2006), we recover 69 of the 77 Lacy et
al. (2007) sources selected for optical follow-up.  Excluded are the
two starburst galaxies, one high-z unidentified galaxy, three Type 2
AGN, one reddened Type 1 AGN, and 1 Type 1 AGN whose slope of $\alpha
= -0.41$ falls just short of our cut of $\alpha \le -0.5$.  The high
reliability of this luminous color-selected sample is therefore
consistent with our findings in \S6 that while non-\plaga\
color-selected AGN are subject to contamination by star-forming
galaxies, sources selected via a power-law criterion are reliable.

\section{IRAC Power-law selection}

\begin{figure}
\epsscale{0.9}
\plotone{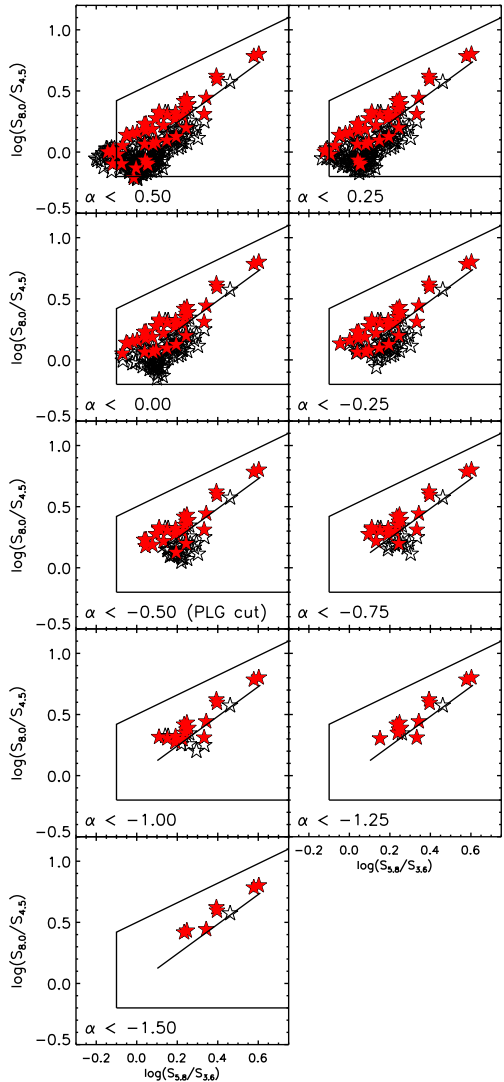}
\caption{Position in Lacy et al. (2004) color space of \plagas\
as a function of the power-law slope cut, $\alpha$.  X-ray--detected
sources are shown as filled (red) symbols.}
\end{figure}

\begin{figure}
\epsscale{0.9}
\plotone{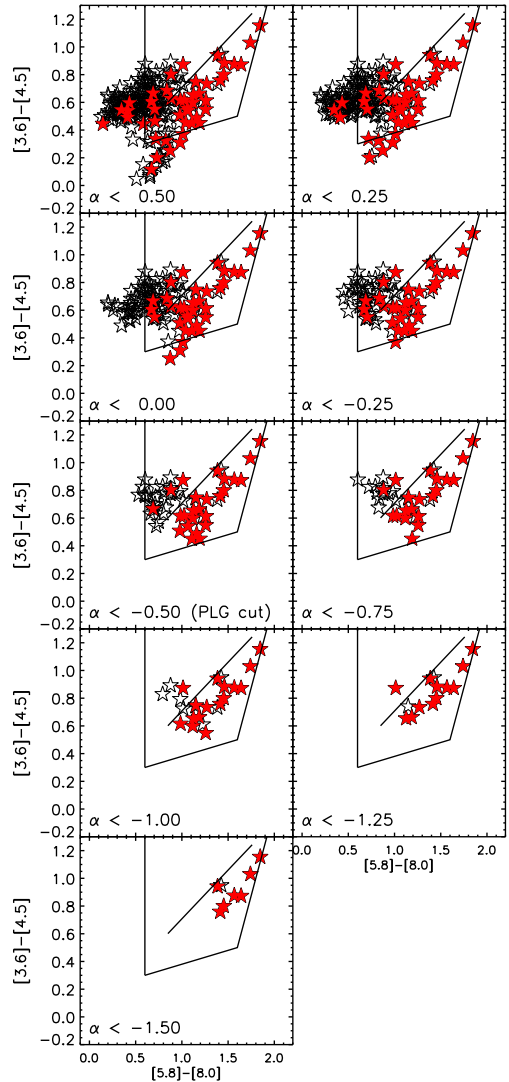}
\caption{Position in Stern et al. (2004) color space of \plagas\
as a function of the power-law slope cut, $\alpha$.  X-ray--detected
sources are shown as filled (red) symbols.}
\end{figure}

Thus far, we have focused primarily on the reliability of the AGN
color selection criteria of Lacy et al. (2004) and Stern et
al. (2005).  In doing so, we have separated out the subset of sources
that meet the power-law criteria of Alonso-Herrero et al. (2006) and
Donley et al. (2007).  Here, we discuss the \plaga\ selection
itself, in particular the consequences of various choices of limiting
power-law slope, $\alpha$.

Our default cut of $\alpha < -0.5$ was chosen to match the spectral
indices of typical AGN (e.g. Alonso-Herrero et al. 2006, Donley et
al. 2007).  In the optical, AGN have spectral slopes of $\alpha = 0.5$
to -2 (SDSS, Ivezi{\'c} et al. 2002), with a mean value of $\alpha
\sim -1$ (Neugebauer et al. 1979, Elvis et al. 1994).  In the
IRAC bands, broad-line AGN exhibit similar slopes, with a mean value
of $\alpha = -1.07 \pm 0.53$ (Stern et al. 2005).

In Figures 7 and 8, we plot \plagas\ selected via cuts in $\alpha$
ranging from +0.5 to -1.5.  At high redshift ($z\sim 1.5-2$) the IRAC
bands trace the blue side of the stellar bump.  It is therefore not
surprising that at the bluest slopes ($\alpha = +0.5$), the \plaga\
sample is dominated by the population of high redshift
X-ray--non-detected star-forming galaxies discussed above. As the
required slope reddens towards our cut of $\alpha < -0.5$, the
high-redshift star-forming galaxies gradually drop out of the sample,
and the X-ray detection fractions rise from 21\% at $\alpha=+0.5$ to
30\%, 45\%, 67\%, 88\%, and 80\% at $\alpha = 0.0, -0.5, -1.0, -1.5,$
and -2.0.  While choosing a redder cut in $\alpha$ therefore increases
the apparent reliability of the \plaga\ selection, it also decreases
the number of galaxies selected, and may exclude interesting heavily
obscured, X-ray--non-detected AGN like those seen in the CDF-N
(e.g. Donley et al. 2007).

The X-ray--non-detected \plagas\ in the current sample tend to be the
faintest sources both in the IRAC bands and at 24~\micron, where their
mean flux density, 146~\microjy, is over a factor of 2 lower than that
of the X-ray--detected sample, 334~\microjy.  The low X-ray detection
fraction of these faint sources may therefore be due simply to their
systematically lower fluxes.  At $z>2.6-2.9$, however, the
star-forming ULIRG templates have IRAC SEDs that meet the \plaga\
criteria, although dropping the power-law slope criterion to $\alpha
\le -1.0$ and $\alpha \le -1.5$ raises this redshift range to
$z>2.74-3.92$ (depending on the template) and to $z>3.4-4.2$,
respectively (with the IRAS~22491-1808 template never reaching an
$\alpha$ of -1.5).  It is therefore possible that the X-ray
non-detected \plagas\ (which tend not to have redshift estimates, to be
faint, and to lie at relatively high $\alpha$) are high-redshift
star-forming galaxies.  To test this hypothesis, we plot in Figure 9
the ratio of the MIPS 24~\micron\ flux density to that of the
3.6~\micron\ IRAC band.  At the redshifts of interest ($z>2.6$), this
flux ratio allows a direct comparison of the hot dust emission at
5-7~\micron\ to the stellar emission at $\sim 1$~\micron.  Overplotted
on the colors of our MIPS-selected sample are the redshifted colors of
the AGN templates of Polletta et al. (2007, see Table 1) and the
purely star-forming LIRG and ULIRG templates of Rieke et al. (2008).
The mean colors of
\plagas\ without redshift estimates are given by large symbols placed
at $z=2.6$, the redshift above which contamination by star-forming
galaxies is possible. The circle, square, and triangle represent all
\plagas\ without redshifts, those that are detected in the X-ray, and
those that are not, respectively.

\begin{figure}
\epsscale{1.1}
\plotone{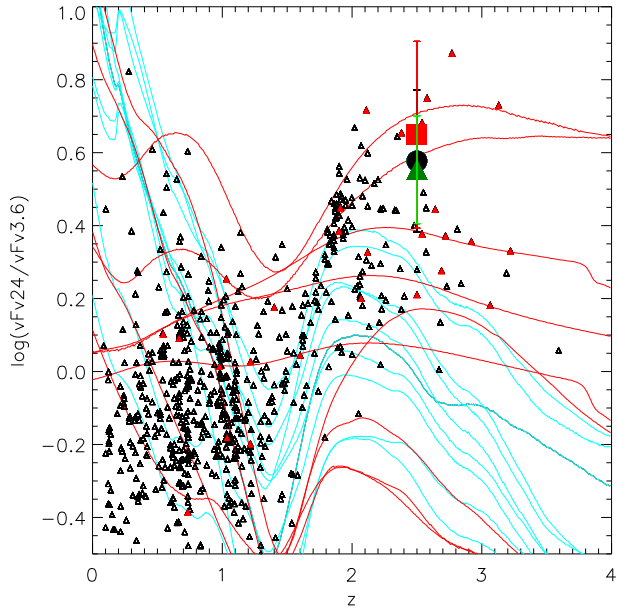}
\caption{Observed 24~\micron\ to 3.6~\micron\ color as a function of redshift.  
The tracks represent the AGN templates of Polletta et al. (2007, red)
(see Table A1), and the purely star-forming LIRG and ULIRG templates of
Rieke et al. (2008, cyan).  Triangles represent all sources in our
sample with redshift estimates, and filled (red) triangles indicate a
\plaga\ source.  The mean colors of \plagas\ without redshift
estimates are given by a large circle (all \plagas), a large square
(X-ray--detected \plagas), and a large triangle (X-ray--non-detected
\plagas).  We arbitrarily placed these mean colors at a redshift of
$z=2.6$, above which contamination of the \plaga\ sample by
star-forming galaxies is possible. The colors of these sources,
however, are consistent with AGN, not star-forming galaxies.}
\end{figure}

As can be seen in Figure 9, the X-ray--detected and
X-ray--non-detected \plagas\ lacking redshifts have consistent
24~\micron\ to 3.6~\micron\ colors, suggesting that there is no
significant difference between these two sub-samples.  In addition,
while the colors of these sources are consistent with those of AGN,
they lie well above those of star-forming galaxies, not only at high
redshift, but at all redshifts greater than $z \sim 0.6$.  In other
words, the \plagas\ that are undetected in the X-ray and that lack redshifts 
(preventing us from testing their reliability as in \S5.3) appear to
have more hot dust emission than can be explained by purely
star-forming templates, especially at high-z where star-forming
contamination of the \plaga\ sample is most likely.  This result
suggests that these sources are more likely to be AGN than
star-forming galaxies.  Other lines of evidence, e.g. variability
(Klesman \& Sarajedini 2007), and X-ray properties (Steffen et
al. 2007), tend to support this conclusion. In addition, we note from
Figure 9 that a number of the other galaxies at $z\sim2$ have colors
slightly redder than those predicted by star-forming galaxies at this
redshift.  This behavior may be due to a minor issue with the
templates, a small AGN contribution to the MIR flux density, or to
reddening that exceeds that seen in our local LIRG/ULIRG templates.

Finally, the segregation of the X-ray--detected and
X-ray--non-detected sources about the power-law locus, seen in Figures
7 and 8, warrants discussion, as this was not seen in the CDF-N
(Donley et al. 2007).  This behavior appears to be due largely to the
different selection methods and limiting fluxes of the two samples.
While the CDF-N \plaga\ sample was selected on the basis of IRAC
fluxes, the current sample was selected from a flux-limited MIPS
sample with far deeper IRAC data.  If we require the CDF-S sample to
meet the IRAC detection limits of the CDF-N study, the
\plaga\ sample size decreases by a factor of 2 and the X-ray detection
fraction rises to 78\%. The segregation in the \plaga\ sample is also
greatly reduced, due largely to the loss of many of the
X-ray--non-detected sources.  This high X-ray detection fraction of
78\%, however, is surprising as it is significantly higher than the
55\% found for the CDF-N \plaga\ sample. To test whether this change
is a result of the much improved IRAC data, we recalculated the
spectral properties of the galaxies in the CDF-N using the most
current IRAC data, which now includes the super-deep GOODS-N coverage
not previously available.  Drawing \plagas\ from the same initial
sample used in Donley et al. (2007), we find an updated X-ray
detection fraction of 57\%, a value nearly identical to that found
previously.  For consistency with the current sample, we further
restricted the \plaga\ sample to those sources detected in the
super-deep GOODS field, where the X-ray exposure is generally higher.
Doing so raises the X-ray detection fraction to 65\%.  The remaining
offset in the X-ray detection fractions of the IRAC flux-limited
\plagas\ in the GOODS-N and GOODS-S fields, 13\%, therefore appears to
be due primarily to cosmic variance and small number statistics.

\section{IR-Excess Selection}

\begin{figure*}
\epsscale{1}
\plottwo{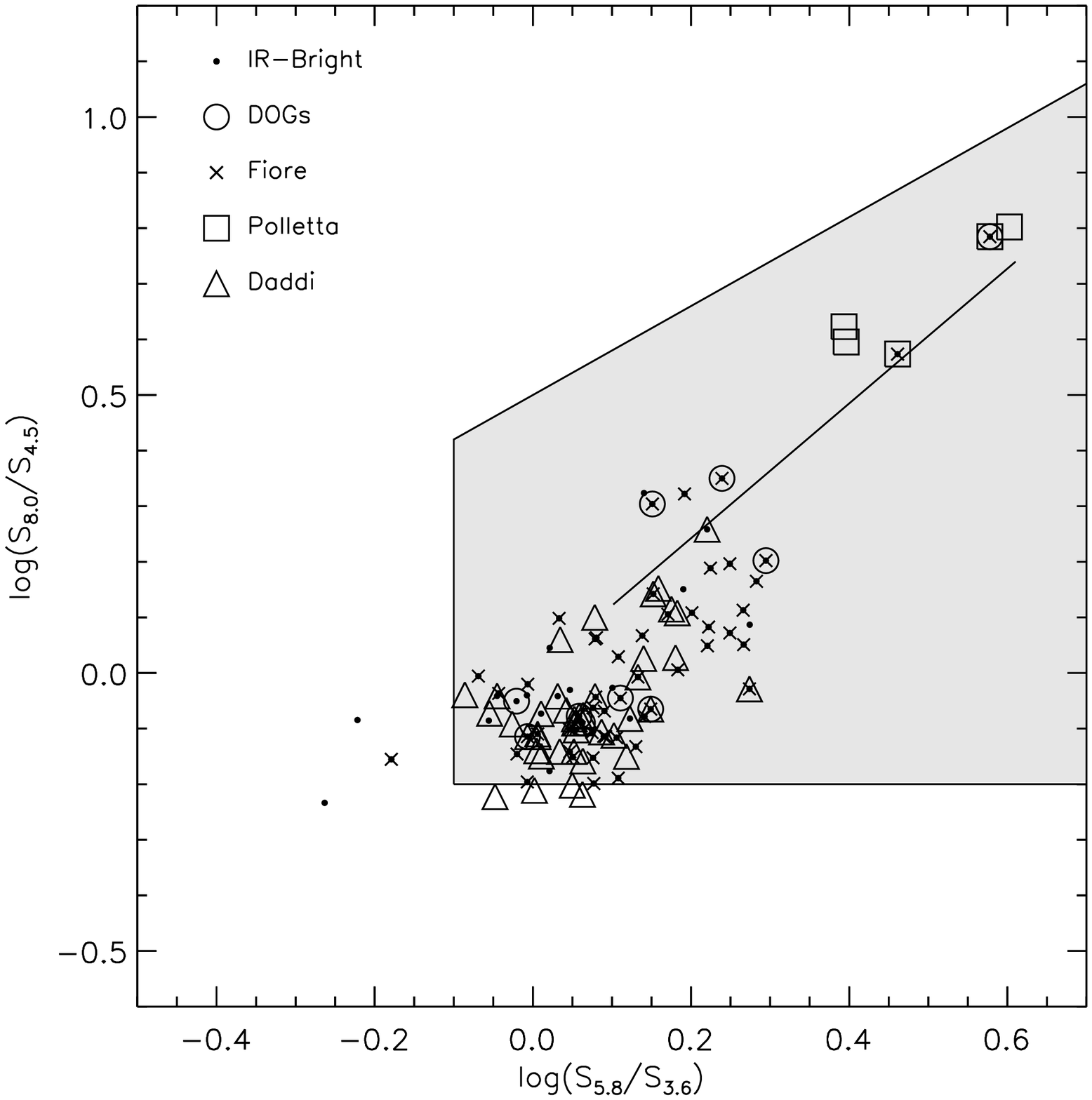}{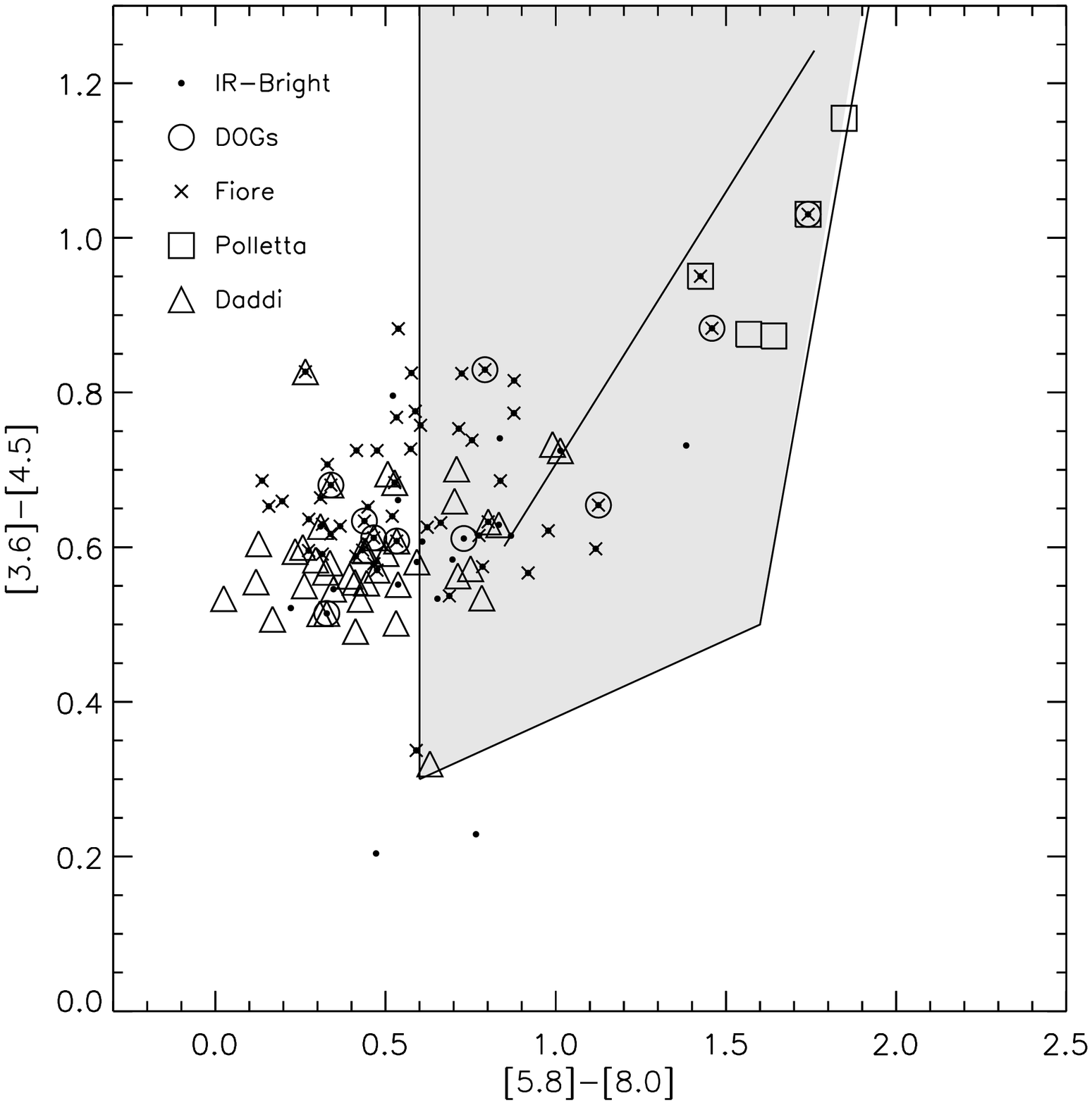}
\caption{Position in Lacy et al. (left) and Stern et al. (right) color-space of the IR-excess
sources.}
\end{figure*}

Both high-redshift star-forming galaxies and AGN can exhibit extremely
red infrared to optical colors when heavily dust-obscured (e.g. Dey et
al. 2008).  While the relative fractions of AGN and star-forming
galaxies among IR-excess samples is still a matter of debate, the
fraction of dust-obscured galaxies (DOGs) with AGN-like power-law SEDs
has been shown to decrease with decreasing flux from $\sim 70\%$ at $f
= 1$~mJy to $\sim 20-25\%$ at $f \le 300$~\microjy\ (Dey et
al. 2008). We might therefore expect a relatively small AGN fraction
amongst the faint IR-excess sources in the GOODs fields, whose mean
24~\micron\ flux is 215~\microjy.  Fiore et al. (2008), however,
estimate that 80\% of their red, IR-excess galaxies are heavily
obscured, Compton-thick AGN.  Here, we probe the nature of the
IR-excess sources and examine the overlap between the IR-excess
galaxies and the X-ray, color-selected, and power-law samples
discussed above.

The positions of the IR-excess galaxies in Lacy and Stern IRAC
color-space are shown in Figure 10, and their overlap with the X-ray,
\plaga, and color-selected samples is given in Table 2.  Of the 101
IR-excess sources, 24 (24\%) are \plagas\ and all but 7 meet either
the Lacy, Stern, or power-law criteria.  At least 1/4 of the IR-excess
galaxies therefore show evidence for AGN-heated dust (with the
fraction rising to 100\% for the luminous AGN selected via the
Polletta et al. (2008) criteria, as expected).  Not all \plagas,
however, have IR-excess colors.  Of the 55 \plagas, only 44\% meet one
or more of the IR-excess criteria. The same can be said for 33\% of
the IRAC color-selected galaxies, but for only 2\% of the IR-normal
galaxies.

Of the IR-excess sources, only 15\% have cataloged X-ray counterparts,
with the fraction rising to 45\% when weak ($\ge 2
\sigma$) X-ray detections are considered (see Table 2).  Of the X-ray--cataloged
sources, 47\% are \plagas, 47\% are color-selected galaxies, and 1
(6\%) is IR-normal.  The statistics are quite different for those
sources with no cataloged X-ray emission, where the power-law fraction
drops to 19\% and the color-selected fraction rises to 73\%.

While a strong argument can be made for the AGN nature of the 32
IR-excess sources with cataloged X-ray counterparts and/or \plaga\
SEDs, what can be said about the remaining IR-excess sources?  As
discussed in \S3, the concentration of the redshifts of these sources
about $z=2$ may be due to the passage of the 7.7~\micron\ aromatic
feature through the 24~\micron\ band, suggesting a significant
contribution from star-formation.  In addition, the vast majority of
color-selected sources at $z=2$ (of which these are a subset) have
IRAC colors consistent with those of star-forming galaxies or
low-luminosity AGN whose IRAC SEDs are dominated by the host galaxy.
In combination with the lack of significant X-ray emission, these
facts suggest that the remaining 68\% of the IR-excess sample could be
either star-forming galaxies, or extremely obscured AGN.  To
distinguish between these two possibilities, we consider in more
detail the sources selected via the Daddi et al. and Fiore et
al. criteria.

\vspace*{0.5cm}

\subsection{Daddi et al. (2007) Compton-thick AGN candidates}

Of the 88 IR-excess AGN candidates selected in the ISAAC region of
GOODS-S by Daddi et al. (2007a), the list of which was kindly provided
by D. Alexander (private communication, 2008), 42 fall in our
MIPS-selected sample.  Twenty-three of the remaining 46 galaxies have
MIPS fluxes that fall below our cut of $f_{\rm 24~\micron} \ge
80$~\microjy, and 6 galaxies lack MIPS and/or IRAC counterparts in our
catalogs, indicating that they too are faint.  Seven galaxies were
excluded because of observable blending in the MIPS or IRAC bands, and
an additional 10 were removed from our sample because of multiple
optical counterparts.

The positions of the Daddi et al. sources in IRAC color-space are
shown in Figure 10.  Thirty-eight of the 42 Daddi et al. sources in our
sample meet the Lacy et al. IRAC color-selection criteria.  Only 10
meet the Stern et al.  AGN selection criterion, and only 5 are
\plagas\ (4 of which have relatively shallow slopes of only $\alpha \sim -0.6$). 
In addition, only 15 (36\%) meet any of the other IR-excess criteria.
This is not surprising, as the MIR emission from the Daddi et
al. IR-excess sources exceeds that of a typical star-forming galaxy by
a factor of only $\gsim 3$ (Daddi et al. 2007b).  Of the 42 IR-excess
galaxies, 18 have IRAC colors that lie $>1 \sigma$ from the
redshift-appropriate star-forming contours in either Lacy or Stern
color-space. Only 3 and 1, however, have colors that lie outside the
$2 \sigma$ and $3 \sigma$ contours, respectively.  As discussed above,
however, this indicates only that any AGN activity can not be
identified on the basis of the MIR IRAC colors alone.

To determine the nature of the Daddi et al. (2007) sources in our
sample, we therefore turn to the X-ray data. While none of the
IR-excess galaxies are individually detected in the hard X-ray band
(by definition), 3 have soft X-ray detections and 13 have faint ($>
2\sigma$) X-ray counterparts.  Twenty remain X-ray--undetected, and 6
sources have a nearby X-ray counterpart that prevents an accurate test
for low-$\sigma$ X-ray flux.  Using the procedure outlined in Steffen
et al. (2007), with the only change being our slightly different
choice of source aperture radius, 2\arcsec, we coadded the three
sources detected in the soft band, whose soft-band luminosities of
$L_{\rm x} > 10^{42}$~ergs~s$^{-1}$ indicate that they are AGN. We
verify that our stacking method reproduces the results of Daddi et
al. (2007b) for the same sample of 59 IR-excess galaxies used in that
work.  While the coaddition of the 3 soft X-ray detected sources did
not lead to a hard band detection, we place a $3\sigma$ limit on the
hard-band flux that constrains their column density to $N_{\rm H}
\lsim 2 \times 10^{22}$~cm$^{-2}$, assuming an intrinsic photon index
of $\Gamma = 1.8$.

A coaddition of the 13 weakly-detected sources leads to a $3.3 \sigma$
hard-band detection, a $9.7 \sigma$ soft-band detection, a hardness
ratio of $HR = -0.31$, and a photon index of $\Gamma = 1.4$.  If we
assume that these sources are AGN at their mean redshift of $z=1.81$,
the hard to soft flux ratio corresponds to an obscured column density
of $N_{\rm H} = 3.6 \times 10^{22}$~cm$^{-2}$.  At this modest
obscured yet Compton-thin column density, the observed soft band flux
is only attenuated by a factor of 2, implying that the sources must
have relatively low luminosities.  Indeed, the rest-frame 2-10 keV
absorption-corrected luminosity derived from the observed soft-band
flux is only $L_{\rm x} = 1.6 \times 10^{42}$~ergs~s$^{-1}$.

However, Daddi et al. (2007b) argue that the coadded soft X-ray flux
of their full IR-excess sample can be attributed to star-formation.
When they subtract this component from the detected fluxes, the
hardness of the remaining X-ray emission implies a significantly
larger (e.g. Compton-thick) column density as well as a larger
absorption-corrected luminosity.

To test the origin of the X-ray emission, we plot in Figure 11 the
2-10 keV luminosity and 25~\micron\ power ($\nu L \nu$) of a sample of
starburst and AGN-dominated galaxies and ULIRGs drawn from
Franceschini et al. (2003), Ranalli et al. (2003), and Persic et
al. (2004). The 25~\micron\ luminosities were extracted from the IRAS
Revised Bright Galaxy Sample (RBGS, Sanders et al. 2003) when
available, and from ISO (Klaas et al. 2001) or the IRAS Faint Source
Catalog otherwise.  As both the hard X-ray and MIR flux densities of
star-forming galaxies trace the current star-formation rate
(e.g. Ranalli et al. 2003, Franceschini et al. 2003), these two
luminosities are well correlated for starbursts and
starburst-dominated ULIRGS (though there is a hint of a turnover to
lower X-ray luminosities amongst the ULIRG sample).  AGN, however,
show an increased X-ray output for their observed 25~\micron\ flux
density.

\begin{figure}
\epsscale{1.0}
\plotone{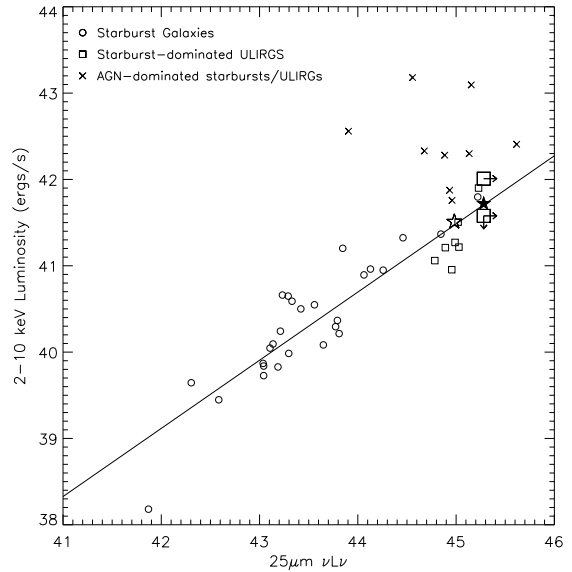}
\caption{Rest-frame 2-10 keV luminosity vs. rest-frame 
25~\micron\ power.  The small symbols represent the starburst and
ULIRG samples of Franceschini et al. (2003), Ranalli et al. (2003),
and Persic et al. (2004), where starburst dominated sources are given
as squares and circles, and AGN-dominated sources are given by
crosses.  The solid line gives the best-fit linear relationship
between the X-ray and 25~\micron\ luminosities of the
star-formation-dominated sources.  The large filled and open stars
represent the full IR-excess and IR-normal samples from Daddi et
al. (2007b), respectively, and the large squares represent the X-ray
weakly- and non-detected members of the Daddi et al. sources in our
sample.  Because these sources comprise the brighter subset of the
full Daddi et al. sample, we treat the full 25~\micron\ flux density
(derived from the coadded 70~\micron\ flux density given by Daddi et
al. (2007b)) as a lower-limit. }
\end{figure}

To compare the Daddi et al. sample to these local starburst and ULIRG
samples, we convert the observed soft-band (rest frame $\sim 1.5-6$
keV) luminosity to a rest-frame 2-10 keV luminosity, and the observed
70~\micron\ (rest-frame $\sim 23$~\micron) flux density given by Daddi
et al. (2007b) to a rest-frame 25~\micron\ power.  The mean
luminosities of the full Daddi et al. (2007b) IR-normal and IR-excess
samples fall on the best-fit correlation for star-forming galaxies,
confirming that their observed soft X-ray flux can be attributed to
star-formation. The X-ray luminosity of the X-ray weakly-detected
IR-excess galaxies in our MIPS sample, however, is somewhat higher
than predicted.  The AGN origin of the soft X-ray emission is
supported by the 2-10 keV luminosity implied by the soft X-ray flux,
$L_{\rm x} = 1.6 \times 10^{42}$~ergs~s$^{-1}$. If the soft X-ray flux
has a significant contribution from the AGN, then it cannot all be
subtracted from the AGN spectrum, resulting in column densities lower
than estimated by Daddi et al. (2007b).

\begin{figure*}
\plottwo{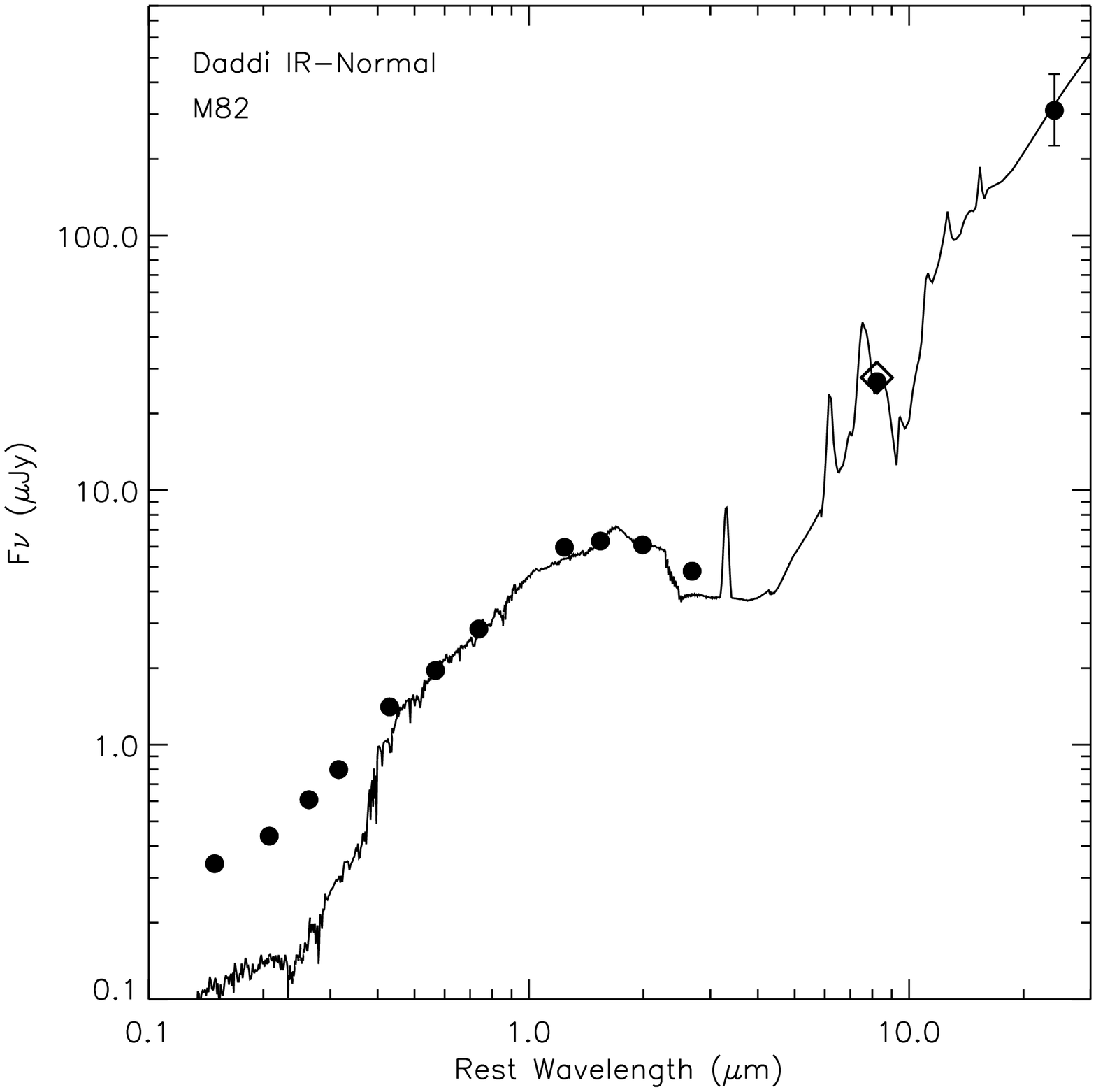}{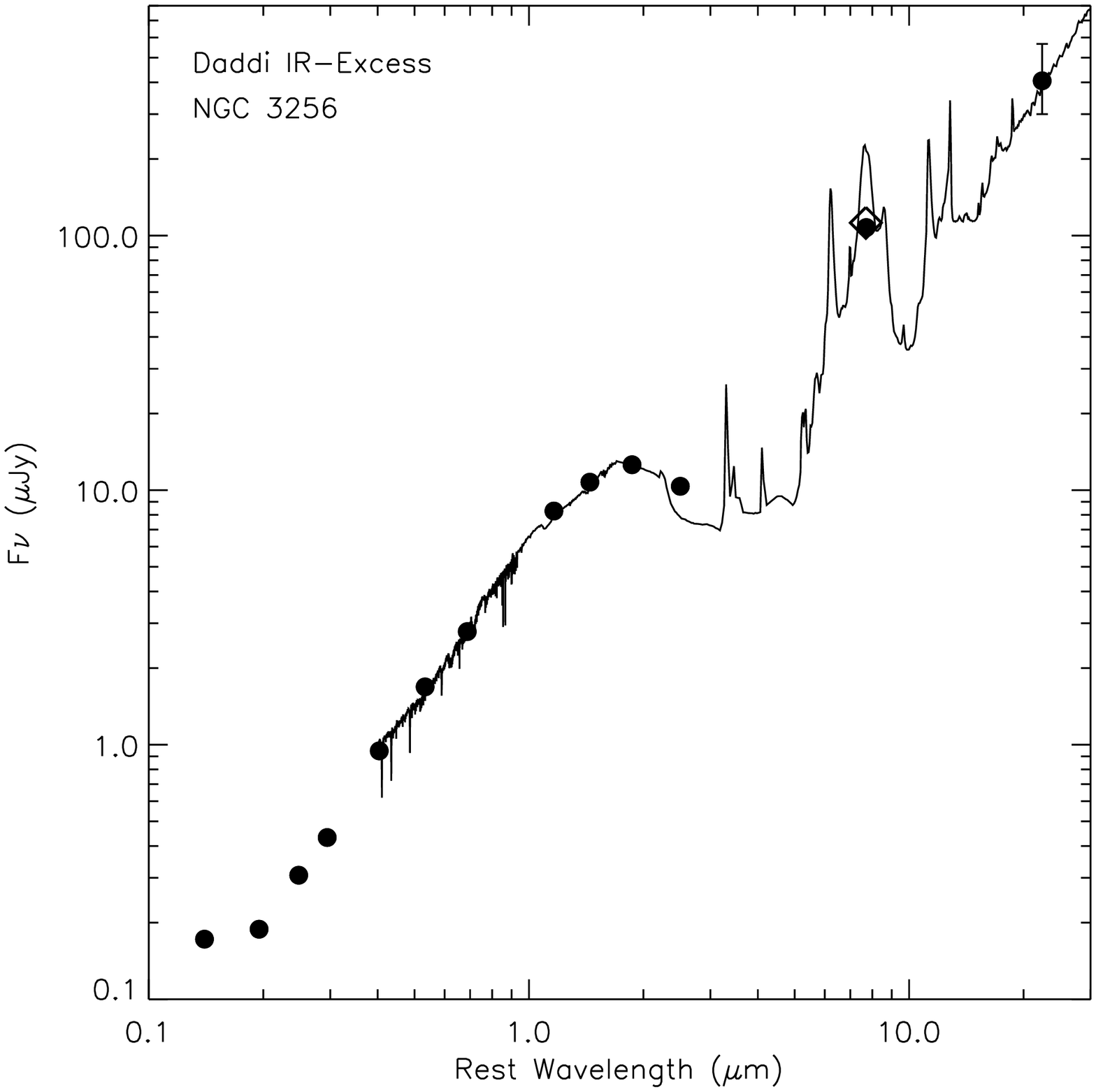}
\caption{Median SEDs of the full Daddi et al. (2007a,b) IR-normal 
and IR-excess GOODS-S samples, as given in Figure 4 of Daddi et
al. (2007b). Overplotted are the \textit{purely star-forming} SEDs of
the starburst galaxy M82 (left) and LIRG NGC 3256 (right) (The Rieke
et al. (2008) templates only extend down to $\sim 0.4$~\micron). We
have applied an additional reddening of $A_{\rm V} = 0.8$ to the
NGC~3256 template. The open diamonds represents the flux that would be
measured in the MIPS 24~\micron\ band when the bandpass is convolved
with the SED.  The optical-MIR SEDs of both the IR-normal and
IR-excess sources are well-fit by star-forming SEDs.  }
\end{figure*}

However, the indicated AGN contribution is modest, and is based on an
average.  Therefore, it is likely that there will be a significant
range of star-formation--corrected absorbing columns, including some
that are Compton thick. Indeed, Alexander et al. (2008) show that a
sample of 6 spectroscopically-confirmed Compton-thick AGN would be
selected via the Daddi et al. (2007) method, although 4 of the 6 have
infrared excesses of a factor of $>100$, and thus represent the most
extreme IR-excess sources.

A coaddition of the 20 X-ray non-detected Daddi et al. sources in our
sample leads only to a marginal $2.0 \sigma$ detection in the full
band (0.5-8 keV), and $1.8 \sigma$ and $1.0 \sigma$ detections in the
hard and soft bands, respectively. A $3 \sigma$ limit on the soft flux
gives a 2-10 keV luminosity of $L_{\rm x} \le 3.8 \times
10^{41}$~ergs~s$^{-1}$, fully consistent with a star-forming origin
(see Figure 11). (The 2-10 keV luminosity derived from the marginal
full-band detection is even lower: $L_{\rm x} \le 2.2 \times
10^{41}$~ergs~s$^{-1}$). The near-detection in the hard-band, however,
suggests that these sources may have relatively hard X-ray spectra.
By coadding the weakly and non-detected sources, the photon index
drops to $\Gamma = 1.04 ^{+0.16}_{-0.14}$, from a value of $\Gamma =
1.39 ^{+0.21}_{-0.17}$ measured for the weakly-detected sources. There
is therefore marginal ($1 \sigma$) evidence that the X-ray
non-detected sources have spectra harder than those of their
weakly-detected counterparts.  Does this hard spectrum confirm that
these sources are AGN, or might there be another explanation?

\subsubsection{A Star-formation Origin?}

While it is well-known that heavily obscured AGN exhibit hard X-ray
spectra, star-forming galaxies can also produce such spectra.  As
discussed in Persic \& Rephaeli (2002) and Persic et al. (2004), the
hard ($>2-10$ keV) X-ray emission of star-forming galaxies is
dominated by low and high-mass X-ray binaries (LMXBs, HMXBs), with the
HMXB fraction increasing from $\sim 20\%$ at starburst luminosities to
$\sim 100\%$ at ULIRG luminosities.  As shown by White, Swank, \& Holt
(1983), HMXBs have X-ray spectra with $\Gamma = 1.2 \pm 0.2$, a cutoff
energy of 20 keV, and an $e$-folding energy of $\sim 12$ keV (see
e.g. Persic \& Rephaeli 2002). At $z=2$, the 0.5-2 keV and 2-8 keV
X-ray bands sample the rest-frame 1.5-6 keV and 6-24 keV X-ray bands,
and therefore should be minimally affected by the power-law cutoff. A
luminous star-forming galaxy undergoing an isolated burst of
star-formation is therefore expected to display a hard X-ray spectrum
of $\Gamma = 1.0-1.4$ over the energies observed in our sample.  X-ray
binary emission has been proposed as an explanation for the hard
($\Gamma \sim 1.0$) spectrum of Arp 220 (Iwasawa et al. 2001), the
spectra of the apparently starburst-dominated ULIRGS of Ptak et
al. (2003) and Teng et al. (2004), whose photon indices tend to lie at
$\Gamma = 1.0-1.5$, and the starburst-dominated ULIRGS of Franceschini
et al. (2003).

While the hard X-ray properties of the IR-excess sources do not
therefore require an AGN origin, can the same be said of their MIR
properties?  To test the nature of these sources, we first plot in
Figure 12 the median SEDs of the full GOODS-S IR-normal and IR-excess
samples, as given in Figure 4 of Daddi et al. (2007b).  Although these
high-redshift sources appear to be highly luminous (see Figure 11),
there are indications that local templates for lower-luminosity
galaxies are appropriate for them, at least in the optical--MIR (Rigby
et al. 2008). It is therefore not surprising that the optical--MIR SED
of the IR-normal galaxies is well-fit by the M82 SB template of
Polletta et al. (2007).  (The discrepancy between the observed SED and
template at shorter wavelengths is likely due to a difference in the
reddening and/or age of the stellar population.)  The IR-excess SED is
also very well fit by the \textit{purely star-forming} template of the
LIRG NGC~3256 (Rieke et al. 2008), to which we have applied an
additional reddening of $A_{\rm V} \sim 0.8$. While this does not rule
out an AGN contribution to the Daddi et al. IR-excess sample, it does
indicate that an AGN need not be present to produce the observed SEDs
of the IR-excess sources, and that their IR-excesses may simply be due
to strong aromatic emission associated with their systematically higher IR
luminosities, as suggested by Daddi et al. (2007a).

If these IR-excess sources are dominated by star-formation, then it
appears either that their IR SFRs have been overestimated or that
their UV-derived SFRs have been underestimated. The former scenario
could be partially attributed to the larger than unity slope between
24~\micron\ luminosity and Pa$\alpha$-derived SFR (Alonso-Herrero et
al. 2006, Calzetti et al. 2007), which indicates that as the SFR
increases, an increasing fraction of the resulting light is emitted in
the MIR.  A simple proportional relationship between SFR and IR
luminosity (e.g. that of Kennicutt 1998) will therefore increasingly
overpredict the SFR for more luminous infrared galaxies.

An underestimate in the UV-derived star-formation rates could be due
to the inherent difficulties in determining accurate UV SFRs for
luminous, heavily obscured galaxies (e.g. Goldader et al. 2002, Buat
et al. 2005).  While UV extinction is known to correlate with
luminosity (see Goldader et al. 2002; Vijh, Witt, \& Gordon 2003; Buat
et al. 2005), Daddi et al. (2007a) find no difference between the
average derived $A_{\rm 1500}$ values of the IR-normal and IR-excess
samples, despite the systematically higher 8~\micron\ luminosities of
the IR-excess sample (see Figure 2 of Daddi et al. 2007b). Under the
assumption that the radio emission arises purely from star-formation,
the radio luminosities of the 16 radio-detected IR-excess galaxies
similarly indicate the UV SFRs have been underestimated by a mean
factor of 6.4, and a median factor of 3.5 (see Figure 12 of Daddi et
al. 2007a). A stack of all IR-excess galaxies, however, leads to radio
SFRs that are consistent with those derived in the UV.

\subsubsection{Summary}

We have divided the 42 Daddi et al. IR-excess galaxies in our
MIPS-selected sample into three subsamples: those that are X-ray
detected (3), those that are weakly-detected in the X-ray (13), and
those that remain undetected in the X-ray down to $2 \sigma$ (20).
(The remaining 6 sources lie too close to an X-ray source to test for
faint emission.) The 3 X-ray detected sources are AGN with $L_{\rm x}
> 10^{42}$~ergs~s$^{-1}$, but have relatively low obscuration ($N_{\rm
H} \le 2 \times 10^{22}$~cm$^{-2}$).

A coaddition of the weakly-detected sources leads to a hard ($\Gamma =
1.4$) X-ray detection.  If these sources are obscured AGN, the
hardness ratio implies a column density of $N_{\rm H} = 3.6 \times
10^{22}$~cm$^{-2}$.  A hard detection, however, could be attributed
either to obscured AGN activity or to star-formation via
HMXBs. Furthermore, the median SED of the full Daddi et al. (2007b)
IR-excess sample is consistent with that of a local star-forming LIRG,
suggesting that the MIR emission alone also cannot be used to rule out
a star-forming origin.  A comparison of the X-ray and MIR luminosities
of these weakly-detected sources does not conclusively distinguish
between a star-forming or AGN origin for the X-ray emission. There are
therefore 3 possible explanations for the members of this sample of
IR-excess galaxies: (1) the sources are Compton-thick AGN whose soft
X-ray emission can be attributed to star-formation and whose hard
X-ray emission comes from the AGN, (2) they are relatively
low-luminosity, Compton-thin AGN, whose soft X-ray emission can not be
entirely attributed to star-formation, or (3) they are star-forming
galaxies whose soft \textit{and} hard X-ray emission arise from
star-formation.  While the derived 2-10 keV X-ray luminosity of $\sim
10^{42}$~ergs~s$^{-1}$ suggests explanation (2), it is not
sufficiently high to definitively rule out the remaining scenarios.
It is likely that no single possibility applies to all 13 galaxies,
but that the possibilities define the range of their behavior.

The X-ray non-detected sources, when coadded alone and with the X-ray
weakly-detected sample, also appear to have a relatively hard X-ray
spectrum.  Their X-ray luminosity, however, is significantly lower
than that of the weakly-detected sources, $L_{\rm x} \le 3.8 \times
10^{41}$~ergs~s$^{-1}$.  If the hard spectrum arises from HMXBs, the
properties of this sample would be consistent with a star-formation
origin.

In summary, while we cannot rule out an obscured AGN origin for the
Daddi et al. IR-excess sources in our MIPS-selected sample, their
properties may also be consistent with a purely star-formation origin
in the majority of cases. Further analysis, taking into account the
additional 1 Ms of X-ray exposure in the CDF-S, is therefore required
to establish their nature.

\subsection{Fiore et al. (2008) Compton-thick AGN candidates}

Fiore et al. (2008) select IR-excess galaxies with the following
properties: $f_{\rm 24~\micron}/f_{R} \ge 1000$ and $R-K > 4.5$.
Using the MUSIC catalogs (see \S2.3), we selected 64 sources that meet
these criteria and that have 24~\micron\ flux densities $>
80$~\microjy.  However, 9 of these sources were removed from our MIPS
sample because of visible blending in the MIPS and/or IRAC bands (see
\S 2), which may have been responsible for their anomalously high IR to
optical flux ratios (unlike Grazian et al. (2006), we do not attempt
to de-blend such sources).  A visual inspection of the remaining
sources led to the removal of 3 additional sources with blended K-band
fluxes, leaving 52 high-quality IR-excess sources.

The first Fiore criterion was designed to select obscured AGN with
large X-ray to optical flux ratios (see their Figure 2) whose column
densities tend to range from $N_{\rm H} = 10^{22}$ to
$10^{23}$~cm$^{-2}$ (see Fiore et al. 2008 and references therein).
The $R-K$ criterion ensures that only extremely red objects (EROs)
fall in the sample.  The AGN amongst ERO samples tend also to be X-ray
obscured with $N_{\rm H} = 10^{22}-10^{24}$~cm$^{-2}$ (Brusa et
al. 2005).  Obscured AGN are therefore likely to be targeted by these
criteria.  However, these two selection criteria are known to identify
both AGN and star-forming galaxies (e.g.  Alexander et al. 2002,
Doherty et al. 2005, Dey et al. 2008). We therefore examine whether
these criteria are sufficiently stringent to exclude the possibility
that the properties of these sources could arise from star formation.

The first of the three main arguments for the Compton-thick AGN nature
of the Fiore et al. sources is the ability of the obscured AGN
template of IRAS 09104+41091 (Pozzi et al. 2007) to reproduce the
extreme colors of these sources, and the comparable inability of the
M82 and Arp~220 star-forming templates to do the same (see their
Figure 3). When fitting SEDs to the sources in their sample, Fiore et
al. find that only 36\% of their X-ray non-detected sources are
best-fit by elliptical, spiral, M82, N6090, or Arp~220 star-forming
templates.  Of the sources in our MIPS-selected sample that meet the
Fiore et al. criteria and have good redshift fits, we find an even
lower fraction of sources best-fit by these templates: 20\%.  However,
as shown in Figure 13, the purely star-forming IRAS~22491 template of
Polletta et al. (2007) satisfies the Fiore et al. criteria when modest
additional reddening ($A_{\rm V} \le 1.2$) is allowed, indicating that
the extreme colors of these sources can be reproduced not only by
obscured AGN, but by highly reddened star-formation as well. Indeed,
when we include this star-forming template, as well as the
star-forming LIRG/ULIRG templates of Rieke et al. (2008), the fraction
of Fiore sources in our sample best-fit by a star-forming template
rises from 20\% to 66\%.

\begin{figure}
\epsscale{1.0}
\plotone{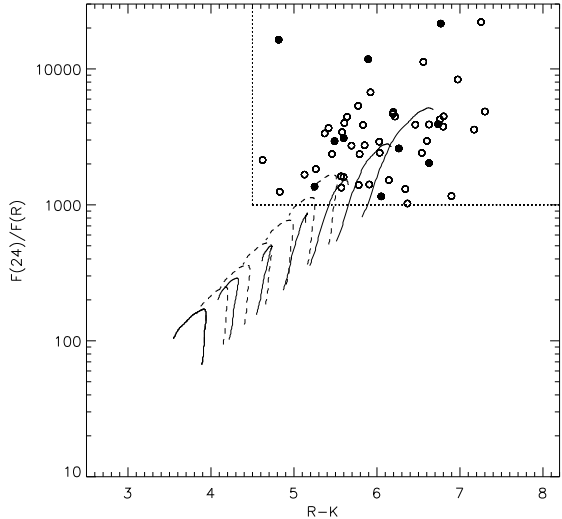}
\caption{Redshifted ($z=1.5-2.5$) tracks of the star-forming template 
IRAS~22491-1808 in Fiore et al. color space.  Additional extinctions
of $A_{\rm V} = $0.0, 0.2, 0.4, 0.6, 0.8, 1.0, and 1.2 were applied
via the SMC (solid, Prevot et al. 1984, Bouchet et al. 1985) or
Calzetti (dashed, Calzetti et al. 2000) extinction laws.  The dotted
lines represent the Fiore et al. (2008) selection criteria, and
circles represent the 52 sources in our sample that meet those
criteria.  The 10 X-ray--detected sources are given as filled
circles. The star-forming template enters the selection region with
only a modest ($A_{\rm V} \sim 1$) amount of additional extinction. }
\end{figure}

The second argument for the Compton-thick nature of the Fiore et
al. sources is the high fraction of heavily obscured AGN ($N_{\rm H} =
10^{23}-10^{26}$~cm$^{-2}$) required to reproduce the hardness ratios
and counts of the stacked X-ray emission.  Using a simulation, Fiore
et al. (2007) estimate that 80\% of their sources are obscured AGN.
The prediction for the X-ray non-cataloged sources in our sample is
somewhat lower: $60\%^{+20\%}_{-40\%}$, although this fraction rises
for the weakly-detected IR-excess sources to $80\%^{+10\%}_{-20\%}$
(F. Fiore, private communication, 2008).

To test this finding for the sources in our sample, we coadded the 36
X-ray non-cataloged Fiore et al. sources that lie far enough from
known X-ray sources to allow an accurate test for faint X-ray
emission.  The coaddition gave a soft detection (0.5-2 keV,
3.9$\sigma$), but only a weak hard detection (2-8 keV,
1.7$\sigma$). The resulting hardness ratio, $HR = (H-S)/(H+S) = -0.17$
is only slightly harder than that of the 10 X-ray cataloged Fiore et
al. sources, $HR = -0.21$, whose column densities fall in the
unobscured to highly obscured (log $N_{\rm H}$~(cm$^{-2}$)$ = 23.9$)
range (Tozzi et al. 2006), with a median value of log $N_{\rm
H}$~(cm$^{-2}$)$ = 22.6$).

If we stack only the 10 sources with weak X-ray counterparts (that
also lie sufficiently far from known X-ray sources), however, we find
a $5.3 \sigma$ hard-band detection, a $4.7 \sigma$ soft-band
detection, and a hardness ratio of $HR = 0.33$, significantly harder
than that of the X-ray cataloged sample. At a redshift of $z=2$, this
$HR$ corresponds to a column density of $N_{\rm H} = 2.7 \times
10^{23}$~cm$^{-2}$.  The observed photon index, $\Gamma = 0.33$,
indicates an obscured AGN origin, as it is inconsistent even with the
moderately hard spectra of HMXB-dominated star-forming galaxies
($\Gamma \sim 1.0-1.4$).  Stacking the remaining 26
X-ray--non-detected Fiore sources does not lead to a detection in any
band.  It therefore appears plausible, at least in this bright subset
of the Fiore et al. sample, that the hard X-ray flux can be attributed
to a small number of obscured, yet mostly Compton-thin, AGN, as
opposed to a large number of obscured, Compton-thick AGN.  

The third argument for the AGN nature of the Fiore et al. sources is
the significant (factor of 30) offset between the IR and UV-derived
SFRs.  However, the inherent difficulties in determining accurate UV
SFRs for luminous, heavily obscured galaxies (e.g. Goldader et
al. 2002, Buat et al. 2005), as well as the systematic uncertainties
of factors of 10-30 in the TIR luminosity (Fiore et al. 2008), make
this the weakest of the 3 arguments.

Of the 52 Fiore et al. sources in our MIPS sample, 10 (19\%) are
therefore X-ray--selected AGN, and 10 (19\%) are weakly detected in
the X-ray, with coadded properties consistent with their being
obscured AGN. As for the remaining 26 sources for which we can test
for faint X-ray emission, we cannot rule out the presence of
Compton-thick AGN.  However, the lack of coadded hard counts from this
X-ray non-detected sample, the significant fraction of such sources
that can be fit by star-forming templates, and the Compton-thin nature
of the X-ray--detected Fiore et al. sources all suggest that many of
these sources, which make up the remaining 56\% of the sample, are
instead star-forming galaxies.

\section{Implications for IR Selection of AGN}

Our evaluation of the performance of the infrared selection methods
allows a preliminary estimate of the overall role of {\it
Spitzer}-discovered AGN in the total population. Of the 109 X-ray
sources in the MIPS-selected sample, 95 have redshifts, and of these,
73 (77\%) have AGN-like X-ray luminosities of (log~$L_{\rm
x}$(ergs~s$^{-1}$)$ > 42$).  We therefore assume a sample of 84 (77\%
of 109) X-ray--selected AGN in the MIPS-selected sample. We now
consider how many IR-selected AGN can be added to this total.

\subsection{Infrared power-law and color-selected AGN}

Of the 55 power-law AGN in our sample, 30 lack cataloged X-ray
counterparts.  If we assume that all of the X-ray weakly- and
non-detected \plagas\ are AGN, the power-law selection criteria
increases the number of known AGN (84) by 36\%.  Of the 25/30 X-ray
non-cataloged sources for which we could test for faint emission, 7
(28\%) show weak X-ray emission at the $>2\sigma$ confidence level,
and 18 show no sign of X-ray emission.  Correcting for the 5 sources
with nearby X-ray counterparts therefore gives an estimate of 8.4
weakly-detected sources.  To place a conservative lower-limit on the
contribution of X-ray non-detected \plagas\ to the AGN population, we
select as AGN those \plagas\ that are either weakly-detected in the
X-ray (8.4) or that have extremely red slopes of $\alpha < -1.0$ (8,
see \S6). Combining these two criteria results in a sample of 11.6 AGN
candidates missed in the X-ray catalogs, for a contribution of 14\%.
We therefore conclude that \plaga\ selection increases the number of
known MIPS-detected AGN by $\sim 14-36\%$.  Further adding the 3
color-selected galaxies that lie outside the 3$\sigma$ star-forming
contours and that lack cataloged X-ray counterparts (after correcting
for the fraction with redshifts and for which the distance from the
star-forming contours could therefore be determined) increases the
contribution of IRAC-selected AGN to $\sim 18-40\%$.

A search for weak X-ray emission from the full sample of
color-selected galaxies results in the detection of 44 additional
sources, 70\% of which have X-ray luminosities typical of AGN (see
Table 2).  Of these 30 AGN candidates, however, 19 (63\%) lie at
$z>1.75$, the redshift above which nearly all (94\%) MIPS sources meet
the Lacy et al. criterion, regardless of their nature
(AGN/star-forming).  Their selection as AGN candidates is therefore
not primarily a function of their IRAC colors, but of their X-ray
properties.  As such, we do not add these additional $\sim 30$ AGN
candidates to our \textit{Spitzer}-selected total.

\subsection{Radio/Infrared-selected AGN}

Radio/infrared selection, in which objects are selected for excess
radio emission relative to that at 24~\micron, provides an alternative
way to identify AGN independently of their optical and X-ray
characteristics.  In Donley et al. (2005), radio-excess AGN are
defined as those sources with log~$f_{24~\micron}/f_{\rm 1.4 GHz} <
0$. Unlike \plagas, radio-excess AGN tend to lie at $z\sim 1$, have
Seyfert-like X-ray luminosities of log~$L_{\rm x}$(ergs~s$^{-1}$)$
\sim 42-43$, and have NIR SEDs dominated by the stellar bump (Donley et
al. 2005).  Of the \plagas\ detected in the CDF-N, only 3\% meet the
radio-excess criteria (Donley et al. 2007).  While there is therefore
almost no overlap between these two AGN populations, their X-ray
detection statistics are very similar: only 40\% of radio-excess AGN
are cataloged in the 2~Ms CDF-N.  If we consider only those
radio-excess AGN with 24~\micron\ flux densities in excess of
80~\microjy, the X-ray detection fraction rises to $\sim 60\%$.  At
this flux limit, the CDF-N sample is complete to radio-excess AGN as
defined by Donley et al. (2005).

In the CDF-N, the X-ray and MIPS--detected radio-excess galaxies (with
X-ray exposures greater than 1~Ms) account for 3\% of the total number
of such sources, and their X-ray non-detected counterparts increase
the number of known AGN by 2\%.  Only 10-15\% of AGN, however, are
radio-intermediate or radio-loud.  This small observed sample of
radio-excess AGN is therefore indicative of an underlying population
at least 7 times larger, which would increase the known population of
MIPS-detected AGN by $\sim 15\%$ if a way could be found to identify
them.

Martinez-Sansigre et al. (2006) also select high-redshift obscured AGN
candidates via a 24~\micron\ and radio flux cut, although of the 21
AGN candidates chosen by them, only 6 (29\%) are sufficiently radio
loud to meet the infrared-to-radio selection criteria used above to
define radio-excess AGN.  Their selection, however, also includes a
3.6~\micron\ IRAC cut designed to select red galaxies.  Because it is
designed for use in shallow surveys, only three galaxies in the GOODS
region of the CDF-N meet their MIPS and radio flux cuts (using the
radio data of Richards 2000), and none are red enough to meet all
three criteria.  This selection method therefore does not contribute
to the AGN sample in the deep CDF-S.

\subsection{IR-excess Galaxies}

As discussed above, the IR-excess samples of Daddi et al. (2007a), Dey
et al. (2008), Polletta et al. (2008), and Fiore et al. (2008) contain
various fractions of AGN and star-forming galaxies, with only the
Polletta et al.  selection criteria unquestionably identifying AGN. Of
the Polletta et al. sources, however, 80\% are detected in the X-ray
catalogs, and the remaining source is a weakly-detected
\plaga, whose contribution to the AGN population has already been
considered.

Daddi et al. (2007a,b) conclude that at least 50\% of their IR-excess
galaxies are Compton-thick AGN. If this hypothesis holds for the
sources in our sample, the Daddi et al. selection criteria would
contribute $\gsim 21$ AGN to the MIPS-selected sample in the ISAAC
field, or $\gsim 31$ AGN to the full MIPS sample (as the ISAAC region
comprises only 68\% of our full survey area).  As discussed in \S7.1,
however, it appears plausible that the properties of many of these
sources can be attributed to star-formation.  We therefore add only
the weakly-detected galaxies, whose mean X-ray luminosity suggests a
likely AGN origin. Of the 13 weakly-detected Daddi et al. IR-excess
galaxies, 1 is a weakly-detected \plaga\ and 1 is a color-selected
galaxy that lies $>3 \sigma$ from the star-forming contours, leaving
11 sources whose contribution is yet to be counted (or 12.6 when we
correct for the sources for which we could not test for weak X-ray
emission).  Further scaling to the full sample region results in an
additional contribution of 18.6 sources, or 22\%.

F. Fiore (private communication, 2008) likewise concludes that $\sim
60\%$ of the X-ray non-cataloged sources in our sample are obscured
AGN.  If so, our MIPS sample in the ISAAC field should contain 25
X-ray weakly or non-detected Fiore-selected AGN.  Of the 36/42 X-ray
non-cataloged Fiore sources in our sample (which do not lie too close
to a known X-ray source to test for faint emission), however, only 10
(28\%) are X-ray weakly-detected (with properties indicative of heavy
obscuration).  The remaining 26 sources show no evidence for X-ray
emission, and have properties that may also be consistent with
star-formation.  We therefore only consider the contribution from the
10 weakly-detected sources.  Of these 10 sources, 3 are
weakly-detected power-law galaxies.  Correcting for the ISAAC field of
view (and the 6 sources for which we could not test for weak X-ray
emission) results in an additional contribution of 12.0 AGN, or 14\%.

\subsubsection{Combined Contribution}

\begin{figure*}
\epsscale{1}
\plottwo{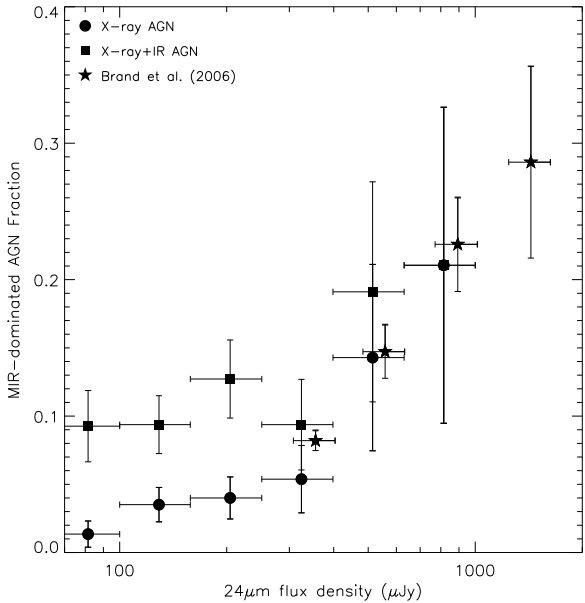}{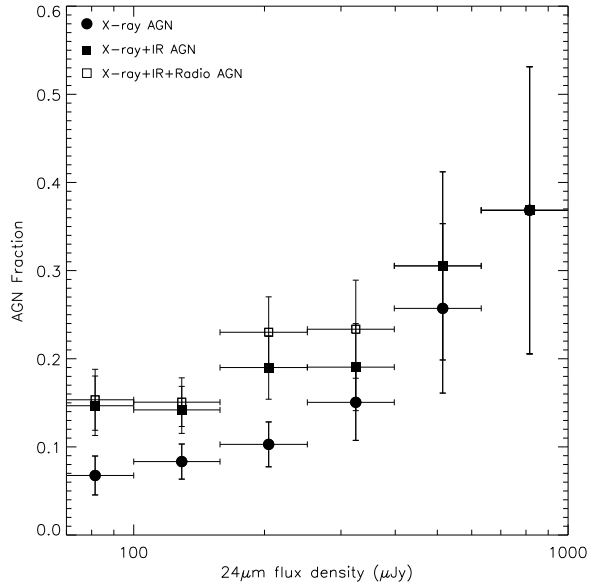}
\caption{Fraction of MIR-dominated AGN, and all AGN (regardless of their 
contribution to the MIR light), as a function of 24~\micron\ flux
density.  The full definitions of 'MIR-dominated' and 'X-ray AGN' are
given in \S9. Error bars represent the $1\sigma$ errors on the source
number counts, and the width of the bins.  The MIR-selected sources
missed in the X-ray comprise the majority of the MIR-dominated AGN at
low flux densities, where the trend towards lower AGN fractions
plateaus.}
\end{figure*}

By combining the contribution of reliable power-law, color-selected,
radio/infrared, and IR-excess AGN candidates, we therefore estimate
that \textit{Spitzer}--selected samples increase the known
X-ray--selected AGN population by $\sim 54-77\%$, down to a
24~\micron\ flux density of 80~\microjy.  In addition, the
radio/infrared selection implies a $\sim 17\%$ contribution from
radio-excess and radio-quiet AGN yet to be identified.  The number of
AGN with 24~\micron\ flux densities $>$ 80~\microjy\ is therefore
$71-94\%$ larger than that of current samples detected both in the
X-ray and at 24~\micron\ in the deepest X-ray fields.

\section{AGN Fraction of the MIR Sample}

We plot in Figure 14a the fraction of the MIPS sample comprised of
MIR-dominated AGN as a function of 24~\micron\ flux density. We
define as 'MIR-dominated' those AGN that (1) meet the
\plaga\ criteria, provided that they are also X-ray--weakly-detected or 
have extremely red slopes of $\alpha < -1.0$ (these criteria were used
above to place a conservative lower limit on the contribution from
\plagas), (2) meet the IRAC color-color cuts of Lacy et al. (2004) or
Stern et al. (2005) and lie outside the $3\sigma$ star-forming
contours, (3) meet the Polletta et al. (2008) criteria, (4) meet the
Fiore et al. (2008) criteria and are X-ray cataloged or
weakly-detected, or (5) meet the Daddi et al. (2007) criteria and are
X-ray cataloged or weakly-detected.  As above, we define as X-ray AGN
those sources with cataloged X-ray counterparts whose total X-ray
luminosities (when redshifts are available) exceed
$10^{42}$~ergs~s$^{-1}$.  Because the \plaga, Daddi et al. (2007), and
Fiore et al. (2008) selection criteria all identify AGN candidates
with mean redshifts of $z\sim2$, the vast majority of the IR-selected
AGN lie at high redshift regardless of their flux density.

As shown by Brand et al. (2006), the fraction of MIR sources dominated
by an AGN drops with decreasing flux density.  We verify this trend,
showing excellent agreement with the Brand et al. results (see Figure
13a), and confirm that it continues down to $\sim 300$~\microjy. At
lower flux densities, however, the fraction of MIR-dominated X-ray AGN
begins to decrease at a much lower rate, and that of all MIR-dominated
AGN plateaus at a value of $\sim 10$\%.  This indicates that a
non-negligible fraction of faint MIR sources are not powered primarily
by star-formation, but by their central engines.  The significant
number of faint 24~\micron\ sources that are AGN-dominated in the MIR
is consistent with the finding of 6/19 such objects in the
spectroscopy of faint 24~\micron\ sources by Rigby et al. (2008).  At
these low flux densities, it is also evident from Figure 14a that the
majority of the MIR-dominated AGN are not detected in the X-ray.  For
instance, at 80~\microjy, the total number of MIR-dominated AGN
outnumbers that of MIR-dominated X-ray AGN by a factor of $\sim 6$.
Using only the X-ray emission as a probe of AGN activity will
therefore result in a serious underestimation of the AGN contribution
to the MIR flux density.

Finally, we plot in Figure 14b the fraction of all AGN in the MIPS
sample (not only those whose MIR flux is dominated by the AGN), as a
function of MIPS 24~\micron\ flux density.  To account for the
contribution of the AGN amongst the 14 X-ray sources lacking
redshifts, we randomly add 11, or 77\% (the AGN fraction of the X-ray
sample), of these 14 X-ray sources to the X-ray AGN sample.  We
further supplement the AGN sample by including the assumed
contribution from radio-excess and radio-quiet AGN, 17\%, or 14
sources.  Because these sources have not been individually identified
in the CDF-S, we draw their flux densities randomly from the range of
observed flux densities in the CDF-N ($80-300$~\microjy).

The X-ray non-detected AGN still comprise the majority of all AGN at
the lowest 24~\micron\ flux densities, and their contribution at
larger flux densities is significant, raising the AGN fraction to
$>15\%$ at all 24~\micron\ flux densities.  As was seen for the
MIR-dominated AGN, the total fraction of AGN rises with increasing MIR
flux density, reaching a value of 37\% at $f_{\rm 24~\micron} \sim
800$~\microjy, 15\% higher than that of MIR-dominated AGN of the same
flux density.  The X-ray AGN fractions we find are somewhat higher
than that of Treister et al. (2006).  While the cause of this offset
is not entirely clear, it is likely to be due at least in part to
cosmic variance, as their 24~\micron\ sample was drawn from GOODS-N,
and ours from GOODS-S.  Once again, the vast majority of X-ray
non-detected AGN at all flux densities lie at $z\sim2$, the mean
redshift of the sources selected via the \plaga, Daddi et al. (2007),
and Fiore et al. (2008) criteria.  Only the radio-selected AGN samples
lie at systematically lower redshifts of $z\sim1$.
 
\section{Summary}

Infrared selection of AGN is a powerful technique.  Using new accurate
star-forming and AGN templates along with a flux-limited MIPS-selected
sample drawn from the GOODS-S field, we critically review three MIR
selection criteria: (1) the IRAC color cuts of Lacy et al. (2004) and
Stern et al. (2005), (2) the power-law galaxy (\plaga) selection
technique of Alonso-Herrero et al. (2006) and Donley et al. (2007),
and (3) the IR-excess selection criteria of Daddi et al. (2007a,b),
Dey et al.  (2008), Fiore et al. (2008), and Polletta et al. (2008).
From this analysis, we then quantify the contribution of
\textit{Spitzer}-selected AGN to the X-ray selected AGN
population. The main conclusions of this paper are as follows:

\begin{itemize}

\item
The majority of non-power-law IRAC color-selected AGN candidates have
IR colors consistent with those of redshift-appropriate star-forming
templates. In comparison, the majority of \plaga\ AGN candidates lie
outside of the star-forming contours.  PLG selection recovers the
majority of high-quality AGN candidates.

\item
The reliability of AGN IRAC color-color selection improves with
increasing flux as high-redshift star-forming galaxies fall out of the
sample.  Nevertheless, the fraction of potential star-forming
contaminants is still high ($\sim 50\%$) at the highest fluxes probed
by our survey ($f_{\rm 24~\micron} \sim 500$~\microjy).

\item
A comparison of the 24~\micron\ to 3.6~\micron\ colors of the X-ray
non-detected \plagas\ to those of AGN and star-forming templates
suggests that the X-ray non-detected \plagas, like their
X-ray--detected counterparts, have more hot dust emission than can be
explained by star-formation alone.

\item
An analysis of the Daddi et al. (2007) IR-excess sources in our MIPS
sample indicates that while these sources may be Compton-thick AGN, it
is also possible that they are low-luminosity, Compton-thin AGN and/or
luminous, highly-reddened star-forming galaxies.

\item
An X-ray stacking analysis of the sources selected via the Fiore et
al. (2008) criteria indicate that $\sim 42\%$ of the sources are
consistent with being obscured AGN, and that the remaining 58\% may be
star-forming galaxies.

\item
Adding secure \textit{Spitzer}-selected power-law, color-color,
radio/IR, and IR-excess AGN candidates to the deepest X-ray
samples directly increases the number of known AGN by $\sim 54-77\%$,
and implies a total increase of $71-94$\%.  This fraction excludes the
full contributions from the Daddi et al. and Fiore et al. AGN
candidates, whose nature is still uncertain.

\item
The fraction of MIR sources dominated by an AGN decreases with
decreasing flux density, but only down to a 24~\micron\ flux density
of $\sim 300$~\microjy.  Below this limit, the AGN fraction levels out
at $\sim 10\%$. This indicates that a non-negligible fraction of faint
24~\micron\ sources are primarily powered not by star-formation, but
by the central engine.  In addition, the majority of AGN at low
24~\micron\ flux densities are missed in the X-ray, indicating that
X-ray emission alone cannot be used to identify AGN, especially
amongst faint IR samples.

\end{itemize}

\vspace*{0cm}

\acknowledgments

We thank M. Polletta for providing templates, M. Dickinson, and
D. Alexander for providing the list of Daddi et al. (2007a,b)
sources. We also thank D. Alexander, F. Fiore, M. Lacy,
D. Stern, and the anonymous referee for discussions and comments that
improved the paper.  Finally, we thank Caltech/JPL for support through
contract 1255094 to the University of
Arizona. P.~G. P.-G. acknowledges support from the Spanish Programa
Nacional de Astronom\'{\i}a y Astrof\'{\i}sica under grant AYA
2006--02358 and AYA 2006--15698--C02--02, and from the Ram\'on y Cajal
Program financed by the Spanish Government and the European Union.

 
\clearpage

\begin{centering}
\begin{deluxetable}{lcccrrrr}
\centering
\tabletypesize{\footnotesize}
\tablewidth{0pt}
\tablecaption {Photometric Redshifts}
\tablehead{
\colhead{Sample}                       &
\colhead{N$_{\rm srcs}$}               &
\colhead{N$_{\rm z}$}                  &
\colhead{N$_{\rm{spec}}$}              &
\colhead{$\Delta$z}                    &
\colhead{$\sigma_{\rm z}$}             &
\colhead{\% with}                      &
\colhead{\% with}                     \\
\colhead{}                             &
\colhead{}                             &
\colhead{}                             &
\colhead{}                             &
\colhead{}                             &
\colhead{}                             &
\colhead{$\Delta z > 0.10$}            &
\colhead{$\Delta z > 0.20$}            
}
\startdata       
All                 & 713  & 649  & 249  & 0.012   & 0.15  & 11.3\%   & 4.6\%  \\
IR-Normal           & 448  & 424  & 187  & 0.012   & 0.12  &  7.2\%   & 2.8\%  \\
Color-Selected      & 210  & 196  &  51  & 0.015   & 0.10  & 20.4\%   & 4.1\%  \\
Power-law           &  55  &  29  &  11  &-0.013   & 0.48  & 40.0\%   & 40.0\% \\
\enddata
\end{deluxetable}
\end{centering}

\begin{deluxetable}{lcccccccc}
\tabletypesize{\footnotesize}
\tablewidth{0pt}
\tablecaption {X-ray Detection Statistics}
\tablehead{
\colhead{}                                   &
\colhead{MIPS\tablenotemark{a}}       &
\colhead{MIPS/ISAAC}       &
\colhead{Daddi et al.}                       &
\colhead{Dey et al.}               &
\colhead{Fiore et al.}             &
\colhead{Polletta et}          &
\colhead{$f_{24~\micron}/f_{\rm R}$}   &
\colhead{All}                             \\
\colhead{}                                   &
\colhead{Sample}                                   &
\colhead{Sample}                                   &
\colhead{(2007)}                                   &
\colhead{(2008)}                                   &
\colhead{(2008)}                                   &
\colhead{al. (2008)}                                   &
\colhead{$> 1000$}                                   &
\colhead{IR-Excess}                                   
}
\startdata       
Total & 713 & 465 & 42 & 10 & 52 & 5 & 71 & 101 \\

\hline
X-rays  & 109 & 76 & 3 & 3 & 10 & 4 & 11 & 15 \\
\hline
Power  & 25 (100\%) & 14  & 0  & 3   & 4   & 4  & 4  & 7   \\
Color  & 35 (79\%)  & 28  & 3  & 0   & 5   & 0  & 6  & 7   \\
Normal & 49 (65\%)  & 34  & 0  & 0   & 1   & 0  & 1  & 1   \\

\hline
Weak X-rays  & 157 & 125  & 13 & 2 & 10 & 1 & 16 & 26  \\
\hline
Power  & 7  (100\%)   & 5   & 1  & 1   & 3    & 1   & 4   & 4  \\
Color  & 44  (70\%)   & 35  & 11 & 1   & 7    & 0   & 11  & 20 \\
Normal & 106 (19\%)   & 85  & 1  & 0   & 0    & 0   & 1   & 2  \\

\hline
No X-rays  & 361 & 226 & 21 & 3 & 26 & 0 & 38 & 51 \\
\hline
Power  & 18   & 12  & 3   & 0   & 7   & 0  & 9  & 11  \\
Color  & 109  & 72  & 15  & 3   & 19  & 0  & 28 & 36  \\
Normal & 234  & 142 & 3   & 0   & 0   & 0  & 1  & 4  \\

\enddata
\tablenotetext{a}{The fraction of X-ray sources with AGN X-ray luminosities of log~$L_{\rm
x}$(ergs~s$^{-1}$)$ > 42$ is given in parentheses.}
\tablecomments{The sum of the weakly and non-detected sources does not equal the full number of X-ray non-cataloged sources. The difference represents the number of sources that
lie too close to a known X-ray counterpart to test for faint X-ray
emission.}
\end{deluxetable}

\begin{centering}
\begin{deluxetable}{lccc}
\centering
\tabletypesize{\footnotesize}
\tablewidth{0pt}
\tablecaption{Fraction of sources that lie outside the star-forming contours}
\tablehead{
\colhead{Selection}               &
\colhead{$1 \sigma$}               &
\colhead{$2 \sigma$}               &
\colhead{$3 \sigma$}               
}
\startdata  

Lacy        & 52\%	        & 21\%       &	10\%       \\ 
Stern       & 40\%	        & 18\%       &	8\%        \\    
Power-law   & 93\%,82\%	        & 79\%,61\%  & 	71\%,43\%  \\    
\hline                  \\
Lacy        & 29\%	        & 12\%       &	5\%       \\ 
Stern       & 22\%	        & 5\%       &	0\%        \\    
Power-law   & 71\%,68\%	        & 61\%,46\%  & 	57\%,39\%  \\

\enddata
\tablecomments{The upper portion of the table assumes no errors on the photometric redshifts.  The lower portion assumes 10\% errors. The two values given for the \plagas\ represent the fraction of sources that lie outside the star-forming contours in Lacy and Stern color-space, respectively.}

\end{deluxetable}
\end{centering}

\begin{centering}
\begin{deluxetable}{llrlrlr}
\centering
\tabletypesize{\footnotesize}
\tablewidth{0pt}
\tablecaption{X-ray detection fraction (and total number) of sources that lie outside the star-forming contours}
\tablehead{
\colhead{Selection}                &
\multicolumn{2}{c}{$1 \sigma$}               &
\multicolumn{2}{c}{$2 \sigma$}               &
\multicolumn{2}{c}{$3 \sigma$}               
}
\startdata  

Lacy              & 21\%   &(19)	        & 26\%  &(10)   &	42\%  &(8)      \\ 
Power-law (Lacy)  & 70\%   &(19)	        & 83\%  &(19)      & 	90\%  &(19)     \\    
Stern             & 19\%   &(5)	                & 25\%  &(3)   &	40\%  &(2)      \\    
Power-law (Stern) & 67\%   &(16)	        & 72\%  &(13)      & 	85\%  &(11)     \\    
\hline                  \\
Lacy              & 29\% &(15)	        & 36\%  &(8)       &	67\%  & (6)     \\ 
Power-law (Lacy)  & 81\% &(17)	        & 89\%  &(16)      & 	88\%  & (15)     \\    
Stern             & 14\% &(2)   	&  0\%  &(0)       &	 0\%  & (0)     \\    
Power-law (Stern) & 80\% &(16)	        & 86\%  &(12)      & 	82\%  & (9)     \\    

\enddata
\tablecomments{The upper portion of the table assumes no errors on the photometric redshifts.  The lower portion assumes 10\% errors.}
\end{deluxetable}
\end{centering}

\begin{centering}
\begin{deluxetable}{ccccccc}
\centering
\tabletypesize{\footnotesize}
\tablewidth{0pt}
\tablecaption{Percent of secure color-selected galaxies vs. 24~\micron\ flux}
\tablehead{
\colhead{24~\micron\ flux cut}      &
\colhead{Lacy}          &
\colhead{Stern}         &
\colhead{Power-law}     \\
\colhead{(\microjy)}      &
\colhead{}              &
\colhead{}              
}
\startdata  
   80   &  52\%  &   40\%  &    93\%,  82\% \\
  100   &  54\%  &   44\%  &    96\%,  84\% \\
  150   &  53\%  &   43\%  &    94\%,  78\% \\
  200   &  60\%  &   63\%  &   100\%,  76\% \\
  300   &  53\%  &   25\%  &   100\%,  72\% \\
  400   &  52\%  &   50\%  &   100\%,  66\% \\
  500   &  56\%  &   50\%  &   100\%,  71\% \\
\tableline
   80   &  29\%  &   21\%  &    72\%,  68\% \\
  100   &  29\%  &   24\%  &    76\%,  72\% \\
  150   &  25\%  &   21\%  &    73\%,  63\% \\
  200   &  32\%  &   27\%  &    92\%,  76\% \\
  300   &  31\%  &   25\%  &    90\%,  72\% \\
  400   &  39\%  &   50\%  &    88\%,  66\% \\
  500   &  50\%  &   50\%  &    85\%,  71\% \\
 
\enddata
\tablecomments{The upper portion of the table assumes no errors on the photometric redshifts.  The lower portion assumes 10\% errors. The two values given for the \plagas\ represent the fraction of sources that lie outside the star-forming contours in Lacy and Stern color-space, respectively.}
\end{deluxetable}
\end{centering}

\appendix

\begin{centering}
\begin{deluxetable}{lll}
\centering
\tabletypesize{\footnotesize}
\tablewidth{0pt}
\tablenum{1}
\tablecaption {A1: Photometric Redshift Templates}
\tablehead{
\colhead{Template}                     &
\colhead{Type}                         &
\colhead{Ref}                  
}
\startdata
Ell2              & 2 Gyr old elliptical     &  (2)    \\
Ell5              & 5 Gyr old elliptical     &  (2)    \\
Ell13             & 13 Gyr old elliptical    &  (2)    \\
S0                & Spiral 0                 &  (1)    \\
Sa                & Spiral a                 &  (1)    \\
Sb                & Spiral b                 &  (1)    \\
Sc                & Spiral c                 &  (1)    \\
Sd                & Spiral d                 &  (1)    \\
Sdm               & Spiral dm                &  (1)    \\
M82               & Starburst                &  (1)    \\
NGC 6090          & LIRG/Starburst           &  (1)    \\
ESO320-G030       & LIRG/Starburst           &  (3)    \\
NGC 1614          & LIRG/Starburst           &  (3)    \\
NGC 2639          & LIRG/Starburst           &  (3)    \\
NGC 3256          & LIRG/Starburst           &  (3)    \\
NGC 4194          & LIRG/Starburst           &  (3)    \\
Arp 220           & ULIRG/Starburst          &  (1)    \\
Arp 220           & ULIRG/Starburst          &  (3)    \\
IRAS 12112+0305   & ULIRG/Starburst          &  (3)    \\
IRAS 14348-1447   & ULIRG/Starburst          &  (3)    \\
IRAS 17208-0014   & ULIRG/Starburst          &  (3)    \\
IRAS 22491-1808   & ULIRG/Starburst          &  (3)    \\
IRAS 22491-1808   & ULIRG/Starburst          &  (1)    \\
\tableline
Mrk 231           & ULIRG/Seyfert 1          &  (1)    \\
Sey1.8            & Seyfert 1.8              &  (1)    \\
TQSO1             & Type 1 QSO               &  (1)    \\
Sey2              & Seyfert 2                &  (1)    \\
NGC 6240          & Starburst/Seyfert 2      &  (1)    \\
IRAS 19254-7245   & ULIRG/Seyfert 2          &  (1)    \\
IRAS 20551-4250   & ULIRG/Buried AGN         &  (1)    \\
QSO2              & Type 2 QSO              &  (1)    \\
\enddata
\tablerefs{(1) Polletta et al. (2007), (2) Silva et al. (1998), (3) Rieke et al. (2008)}
\end{deluxetable}
\end{centering}

\section{Appendix A: Photometric Redshift Techniques}

\noindent We used two methods to determine photometric redshifts. The
first utilizes the extensive high-resolution template set of
P\'erez-Gonz\'alez et al. (2008), which was created by fitting stellar
population synthesis and dust emission models to the $\sim$1500
galaxies in the CDF-N and CDF-S with secure spectroscopic redshifts.
When applied to all spectroscopically-detected IRAC-selected galaxies
in the GOODS-N and GOODS-S, this template library returns photometric
redshifts with $\Delta(z)$$<$0.1 for 88\% of the sources, and
$\Delta(z)$$<$0.2 for 96\%, where $\Delta(z) = (z_p - z_s) / (1+z_s)$
( P\'erez-Gonz\'alez et al. 2008). Because this method relies on
star-forming templates, however, we do not expect it to provide
equally reliable photometric redshifts for galaxies in which the
stellar features are dominated by emission from an AGN.

Our second method is based on the chi-squared minimization routine
{\sc HYPERZ} (Bolzonella, Miralles, \& Pell{\'o} 2000). With a suite
of normal star-forming, LIRG/ULIRG, \textit{and} AGN templates, we can
better account for the range of sources expected in our 24~\micron\
selected sample, albeit with a smaller template library.  To create
this library, we started with a sample of 10 star-forming templates
and 8 AGN templates from Silva et al. (1998) and Polletta et
al. (2007).  The Polletta et al. AGN templates cover a range of
intrinsic obscurations (Type 1 and Type 2) and luminosities (Sey/QSO).
We then supplemented this sample with 10 empirical star-forming LIRG
and ULIRG templates.  The full template sample is listed in Table
1. The LIRG and ULIRG templates, described in detail in Rieke et
al. (2008, in prep.), significantly improve upon previous
semi-empirical templates by constraining the SEDs between 1 and
6~\micron\ with 2MASS and IRAC photometry and by basing the SEDs from
5 to 35~\micron\ on IRS spectra.

Using this template library, we ran HYPERZ on all robust optical--MIR
data for the IR-normal and color-selected galaxies. For the \plagas,
we removed all photometry longward of 3.6~\micron\ (IRAC channel
1). Including the full \textit{Spitzer} photometry for \plagas\
provided little or no additional constraint on potential redshifts,
and limited severely the SEDs for which a good fit could be found. We
allowed $z$ to vary from 0 to 4 and $A_v$ to vary from 0 to 1.2.
While its shape is still relatively unconstrained, a number of studies
have suggested that the extinction curve of AGN most closely resembles
that of the SMC (e.g. Richards et al. 2003).  We therefore assumed by
default the SMC extinction curve of Bouchet et al. (1985) for the
\plagas\ and color-selected galaxies.  For the IR-normal galaxies, we
assume the Calzetti (2000) extinction curve, as this curve was modeled
to represent the extinction properties of starburst galaxies.  To
prevent unrealistic redshift solutions, we applied the redshift- and
model-dependent absolute magnitude cuts of Polletta et al. (2007).
Finally, to increase the weight of the IRAC photometry, which solely
defines the red slope of the stellar bump, we set the errors on the
IRAC photometry to the measurement errors, as opposed to the total
photometric errors.  This procedure is acceptable because many types
of photometric error will have similar effects on the four IRAC bands,
so we are making use of the overall internal consistency expected for
the data.

To improve the likelihood of obtaining a good fit for each source, we
then varied a number of these assumptions, and examined by eye the
resulting fits to choose the most convincing redshift solution.  This
visual inspection is an important characteristic of our work, and was
made possible by the relatively small number of sources in our sample.
First, each \plaga\ and color-selected galaxy was fit by both the SMC
and Calzetti extinction laws.  In most cases, the resulting redshift
fits and solutions varied only slightly (in which case we chose the
SMC-derived fit), but for some galaxies, one of the two extinction
laws provided a clearly superior fit as determined by a visual
inspection. Second, if the absolute magnitude of the resulting
best-fit template did not meet the redshift-dependent absolute
magnitude cut of Polletta et al. (2007), we allowed $M_{\rm B}$ to
vary between -23.7 to -17 for star-forming templates and -28.8 to -19
for AGN templates (e.g. Polletta et al. 2007).  Third, we allowed the
IRAC errors to increase to their total estimated values by adding a
10\% error to the flux; in only 4 cases did this lead to a better fit.
Finally, for the \plagas, we explored fits that did not include the
3.6~\micron\ IRAC channel; only 2 sources benefited from this change.

The final step in our redshift estimation was an independent review of
the redshifts by two authors. By examining by eye the resulting
redshift fits, we chose for each \plaga\ and color-selected galaxy the
best HYPERZ redshift. We then compared this to the redshift fit from
the P\'erez-Gonz\'alez et al. (2008) technique, and replaced the
former with the latter if it clearly provided a better fit. For the
IR-normal galaxies, we use the P\'erez-Gonz\'alez et al. (2008)
redshifts by default, as these are optimized for normal galaxies. We
do, however, remove unconvincing redshift fits, and substitute a solid
HYPERZ redshift if available.  For all sources, we only assign a
photometric redshift if a convincing fit exists.  We rejected 76
sources with poor data or other problems that compromise our redshift
determination.  The photometric redshifts classified as being of high
quality are summarized in Table 1 and shown in Figure 1.

\section{Appendix B: Redshift-dependent Color Selection}

\subsubsection{$z=0-0.25$}

The lowest redshift bin contains no \plagas\
and 2 color-selected galaxies, both of which are detected in the
X-ray, but with low observed X-ray luminosities of log~$L_{\rm
x}$(ergs~s$^{-1}$)$ = 41.3$ and log~$L_{\rm x}$(ergs~s$^{-1}$)$ =
40.9$, indicative of powerful starbursts or very low-luminosity AGN.

\subsubsection{$z=0.25-0.75$}

The second redshift bin contains 3 \plagas, 27 Lacy-selected galaxies,
and 4 Stern-selected galaxies. Of the Lacy and Stern-selected sources,
only 41\% and 25\% are detected in the X-ray, respectively, compared
to 100\% of the \plagas. The average observed 0.5-8 keV X-ray
luminosity of the X-ray--detected color-selected galaxies, log~$L_{\rm
x}$(ergs~s$^{-1}$)$ = 43.0$, is consistent with AGN activity, but is
an order of magnitude less than that of the \plagas\ in this redshift
bin, log~$L_{\rm x}$(ergs~s$^{-1}$)$ = 44.1$.  This is not surprising,
as power-law selection preferentially identifies the most luminous AGN
where the emission from the central engine is able to overpower that
of the host galaxy (Donley et al. 2007).  The AGN that lie amongst the
color-selected sample clearly represent a less luminous AGN population
than those selected via the \plaga\ criteria.  Furthermore, of the 27
Lacy-selected (and 4 Stern-selected) galaxies, only 11 (2) lie outside
the $1\sigma$ star-forming contours in their respective color-space,
with the number dropping to 3 and 1 (2 and 0) at $2\sigma$ and
$3\sigma$.  This suggests that the majority of the color-selected
sources in this redshift bin have infrared colors indicative of
IR-normal galaxies.  While all of the \plagas\ lie outside even the
$3\sigma$ contours in the Lacy color-space, they lie inside the
contours in Stern-space, due primarily to the Sdm template of Polletta
et al. (2007), whose strong 3.3~\micron\ aromatic feature passes
through the 4.5~\micron\ IRAC band at $z \sim 0.4$, causing the
template to enter the Stern AGN selection region at $z=0.23-0.51$.  Of
the 2 \plagas\ whose colors overlap with this template, both have
spectroscopic redshifts of $z>0.67$, and would therefore be 'safe' if
we considered smaller redshift bins.  Removing this template, however,
would have no effect on the number of Lacy or Stern-selected sources
that lie outside the $1\sigma$ contours.

\subsubsection{$z=0.75-1.25$}

We find a higher proportion of secure color-selected galaxies in the
$z=0.75-1.25$ bin, with 10/12 of the Lacy-selected sources and 6/13 of
the Stern-selected sources lying outside the $1\sigma$ star-forming
contours.  This is not surprising, as it is at $z \sim 1$ that the
star-forming contours are best separated from the AGN selection
region. If we extend our test to $2\sigma$, however, the numbers drop
significantly, to 5/12 and 1/13.  The majority of the Stern-selected
galaxies in this redshift bin fill the lower-left corner of the AGN
selection region, the region populated by the Stern-only sources whose
Lacy colors place them in the star-forming locus of color-space.
These star-forming galaxies are likely responsible for the fact that
this is the only redshift bin in which the Stern-selected sources
outnumber the Lacy-selected sources.  Of the 5 \plagas\ in this
redshift bin, all lie outside the $1 \sigma$ star-forming contours in
both Lacy and Stern color-space, and while only 3 lie outside of the
$2\sigma$ contours in Stern color-space, all lie outside of the
$2\sigma$ and $3\sigma$ colors in Lacy color-space.

\subsubsection{$z=1.25-1.75$}

The number of color-selected galaxies rises significantly in the
$z=1.25-1.75$ redshift bin.  The fraction of potential IR-selected
AGN, however, is by far the lowest at these redshifts, with only 6/38
Lacy-selected sources and 3/31 Stern-selected sources falling outside
the $1\sigma$ star-forming contours.  In addition, only 3/38 Lacy
sources and 4/31 Stern sources have X-ray counterparts, further
suggesting that nearly all of the color-selected sources at this
redshift are star-forming galaxies, and not AGN. In contrast, the 2
\plagas\ found in this redshift bin fall outside of the
$1\sigma$ contours in both the Lacy and Stern plots, and both are
detected in the X-ray.

\subsubsection{$z=1.75-2.25$}

The largest number of Lacy-selected sources, 77, is found in the
$z=1.75-2.25$ bin. In addition, an extraordinarily high fraction,
50/77, lie outside of the $1\sigma$ star-forming contours, a
surprising fact given the X-ray detection fraction of only 9\%.  Have
we discovered a significant population of $z\sim2$ obscured AGN
similar to those claimed by Daddi et al. (2007), or is there another
explanation for this population?

The AGN contours (shown in green) provide a better match to the IR
colors of these color-selected galaxies than do the star-forming
templates (shown in blue).  At this redshift, the primary difference
between the IRAC regions of the star-forming and the low-luminosity
AGN templates is the strength of the CO index, which is stronger in
the star-forming galaxies than in the AGN.  The MIPS--selected sources
therefore may have smaller CO indices than those of our local
templates.  One explanation for this offset is that an underlying AGN
continuum has diluted this feature. A lower CO index, however, could
also be attributed to evolution in the metallicity of the
LIRGS/ULIRGS.  At 1/3 solar metallicity, the CO index drops by 4-6
percentage points (McGregor 1987), causing the contours to shift
upwards in Lacy color space by $\sim 0.08$.  While this lessens this
offset between the colors of star-forming contours and the observed
galaxies, it cannot fully account for the observed discrepancy.  Thus,
if a change in the CO index is invoked for the offset, it is likely
that AGN continua are also present.

However, other possible explanations exist.  
If we incorporate into the contours a 10\% error in (1+z) (recall that
the measured $\sigma$ for our photometric redshift fits was 0.15 and
that 11\% of the sources in our spectroscopic redshift sample have
photometric errors $> 10\%$) the resulting contours are far-better
matched to the IR-colors of the color-selected galaxies.  Of the Lacy-
and Stern-selected galaxies, only 17\% and 9\% now lie outside of the
$1\sigma$ contours, with the numbers dropping to 8\% and 0\% at
$2\sigma$ and 4\% and 0\% at $3\sigma$, respectively.

The templates with which the photometric redshifts were best fit can
provide further insight into the sources in this redshift bin. Of the
90 non-\plagas\ in the $z\sim 2$ bin, 75 (86\%) are best fit by a
star-forming ULIRG template, 9 (10\%) are best fit by a ULIRG/hidden
AGN template, 1 (1\%) is best-fit by a type 2 AGN template, and 1
(1\%) is best fit by a type 1 AGN template.  In contrast, all of the
\plagas\ in this redshift range are best-fit by a type 2 AGN template.
Therefore, while we can not rule out the possibility that we have
detected a sample of high-redshift, heavily obscured AGN, the
extremely low X-ray detection fraction of 9\%, the much improved fit
of the contours for which 10\% errors in the photometric redshifts
were included, and the overwhelming fraction of sources for which a
purely star-forming ULIRG provided the best fit to the SED suggest
that it is more likely that the vast majority of color-selected
galaxies in this redshift bin are star-forming galaxies, not AGN.

\subsubsection{$z>2.25$}

The number of color-selected galaxies in the remaining two redshift
bins is relatively low: 18 Lacy-selected sources and 5 Stern selected
sources.  Of these, a large fraction (61\% and 80\%, respectively) lie
outside of the star-forming contours, which cover a comparatively
small portion of the color-space.  At $2\sigma$ and $3\sigma$, the
fractions drop to 39\% and 11\% (Lacy) and 20\% and 0\% (Stern).  At
these redshifts, the X-ray detection fractions of both the
color-selected sources and the \plagas\ are low: 17\% for
the Lacy sources, 40\% for the Stern sources, and 23\% for the
\plagas.  At $z=2.5$, an unobscured AGN with $\Gamma
= 2$ requires a rest-frame 0.5-8 keV luminosity of log~$L_{\rm
x}$(ergs~s$^{-1}$)$ = 42.8$ to meet the flux limit within 1$^{\prime}$
of the CDF-S aimpoint.  At $z=3$, the required value rises to
log~$L_{\rm x}$(ergs~s$^{-1}$)$ = 43.0$, suggesting that if these
sources are AGN, they must have low luminosities, or high obscuring
columns.

\subsubsection{$z={\rm unknown}$}

Not all sources in our sample have redshift estimates. The redshift
completeness for the Lacy and Stern color-selected samples is high
(93\% and 90\%, respectively), but we have high-quality redshifts for
only 49\% of the \plagas.  The difficulty in fitting redshifts to
these sources stems largely from their faint fluxes: the power-law
(color-selected) galaxies without redshifts are significantly fainter,
$V=26.2$ ($25.3$), than those with redshift estimates, $V=25.4$
($V=24.9$), suggesting that these sources may preferentially lie at
high redshift.

The last panels of Figures 5 and 6 show the IRAC colors of
sources without redshift estimates.  Overplotted are the star-forming
contours for $z=0-4$.  Of the sources without redshifts, all but 1 lie
inside the Lacy selection region and all but 9 lie inside the Stern
selection region.  This is not surprising, as sources with non-stellar
continua are the hardest sources to fit.  Of the 13 Lacy-selected
sources, 5 lie outside the $1\sigma$ star-forming contours, and 2 have
X-ray counterparts.  Of the 26 \plagas, 11 lie outside the $1\sigma$
star-forming contours, and 6 are X-ray--detected.  There is a
noticeable concentration of X-ray non-detected \plagas\ towards the
red end of the power-law locus.  These sources are discussed in
more detail in \S6.

\end{document}